\documentclass[journal,11pt,onecolumn,draftclsnofoot,]{IEEEtran}

\usepackage{amssymb}
\usepackage{amsthm}

\usepackage{tikz}
\usetikzlibrary{shapes,arrows}

\usepackage{graphicx}
\usepackage{float}
\usepackage{amsmath}
\usepackage{url}

\usepackage[noadjust]{cite}
\usepackage{caption}

\begin{document}



\title{Distance Estimation Methods for a Practical Macroscale Molecular Communication System}

\author{Fatih~Gulec, Baris~Atakan ~
	\thanks{F. Gulec and B. Atakan are with the Department of Electrical and Electronics Engineering, Izmir Institute of Technology, 35430, Urla, Izmir, Turkey (e-mail: fatihgulec@iyte.edu.tr, barisatakan@iyte.edu.tr).} 
}%

\maketitle 

%
\vspace{-1cm}
\begin{abstract}
Accurate estimation of the distance between the transmitter (TX) and the receiver (RX) in molecular communication (MC) systems can provide faster and more reliable communication. In addition, distance information can be used in determining the location of the molecular source in practical applications such as monitoring environmental pollution. Existing theoretical models in the literature are not suitable for distance estimation in a practical scenario. Furthermore, deriving an  analytical model is not easy due to effects such as boundary conditions in the diffusion process, the initial velocity of the molecules and unsteady flows. Therefore, five different practical methods comprising three novel data analysis based methods and two supervised machine learning (ML) methods, Multivariate Linear Regression (MLR) and Neural Network Regression (NNR), are proposed for distance estimation at the RX side. In order to apply the ML methods, a macroscale practical MC system, which consists of an electric sprayer without a fan, alcohol molecules, an alcohol sensor and a microcontroller, is established, and the received signals are recorded. A feature extraction algorithm is proposed to utilize the measured signals as the inputs in ML methods. The numerical results show that the ML methods outperform the data analysis based methods in the root mean square error sense with the cost of complexity. The nearly equal performance of MLR and NNR shows that the input features such as peak time, peak concentration and the energy of the received signal have a highly linear relation with the distance. Moreover, the peak time based estimation, which is one of the proposed data analysis based methods, yields better results with respect to the other proposed four methods, as the distance increases. Given the experimental data and fluid dynamics theory, a possible trajectory of the molecules between the TX and RX is given. Our findings show that  distance estimation performance is jointly affected  by  unsteady flows and the non-linearity of the sensor. According to our findings based on fluid dynamics, it is evaluated that fluid dynamics should be taken into account for more accurate parameter estimation in practical macroscale MC systems.


\end{abstract}

\begin{IEEEkeywords}
	Molecular Communication, Distance Estimation, Molecular Signal Processing.
\end{IEEEkeywords}

\IEEEpeerreviewmaketitle


\section{Introduction}
Molecular communication (MC) is a promising multiscale communication paradigm which uses molecular signals to transmit and receive information symbols. Biological cells use MC in microscale for several reasons such as finding a nutritional source or coordinating their activities as a swarm or network. It is essential to share information among the cells via MC to form a network \cite{atakan2016molecular}. The developing bio/nanotechnology enables the production of micro and nanoscale devices that mimic living cells, which are called nanomachines (NMs). What makes MC prominent is that it performs complex tasks by forming a nanonetwork, which is established by the interconnection of the NMs \cite{akyildiz2008nanonetworks}, \cite{akyildiz2011nanonetworks}. These nanonetworks can be used in the human body for applications such as targeted drug delivery and immune system support \cite{atakan2012body}.

MC can also be employed in macroscale (cm to m). In nature, animals like bees, flies or ants use MC to send messages over several meters with the help of the pheromones. Similar to the existing methods in nature, pheromones can be exploited for long range communication among NMs as proposed in \cite{gine2009molecular}. On the other hand, the macroscale MC concept was employed in a practical MC system, which consists of an electric sprayer controlled by a microcontroller as the transmitter (TX), a metal-oxide alcohol sensor connected to a microcontroller as the receiver (RX) and alcohol molecules as the messenger molecules, to send information symbols over a few meters \cite{farsad2013tabletop}. The channel and noise models are proposed for this macroscale MC system in \cite{farsad2014channel} and \cite{kim2015universal}. It was observed that there is a nonlinearity in the channel which differs from conventional communication systems. A similar MC system is proposed to be employed in confined structural environments consisting of several pipe types. It was shown that MC provides reliable communication, whereas electromagnetic wave based communication cannot \cite{qiu2014molecular}. The data rate of the practical MC system can be increased by  multiple input multiple output (MIMO) MC system configuration as proposed and experimentally shown in \cite{lee2015molecular} and \cite{koo2016molecular}. In these two papers, an air compressor is used instead of a fan to emit molecules. In \cite{zhai2018anti}, a method to mitigate inter-symbol interference is proposed for an experimental MC system similar to the platform given in \cite{farsad2013tabletop}, except the fan behind the TX. \cite{mcguiness2018parameter} proposes to employ an odor generator as the TX and a mass spectrometer with a quadrupole mass analyzer as the RX for macroscale MC.

The macroscale experimental studies focus mostly on channel modeling which does not agree with theoretical studies in the literature. However, except the propagation delay, these channel models do not clearly state the relationship among the physical parameters such as the distance between the TX and RX or the number of released molecules by the TX. By knowing the distance before the communication starts, a higher communication rate can be achieved via arranging the parameters such as the number of released molecules \cite{atakan2007information, eckford2007achievable, nakano2013transmission, pierobon2013capacity}. Furthermore, the deployment of the receivers can be arranged autonomously in a nanonetwork with the distance information known by the RX. The distance estimation can also be used to find the location of a molecular source in various environmental monitoring applications. Therefore, it is important to estimate the distance between the TX and RX, which is the main purpose of this paper.

The first proposed distance estimation protocols are based on two-way transmission. In \cite{moore2010measuring}, four distance estimation protocols are proposed which are based on measuring the round trip time (RTT) and signal fading in amplitude or frequency for a 1-D diffusion channel. In these protocols, the TX transmits a signal and the RX transmits a feedback signal with a different type of molecule, when the transmitted signal is received. Then, the TX estimates the distance, when it receives the feedback signal. These protocols are expanded in \cite{moore2012measuring}. The RTT protocol is improved for a 2-D diffusion channel with a more realistic microscale MC model in \cite{moore2012comparing}. Since using feedback signals is time consuming and increases interference in the communication channel, it is reasonable to estimate the distance at the RX side with a single transmission. Accordingly, two distance estimation schemes with a one-way transmission are proposed in \cite{huang2013distance}. Here, the RX measures the received peak concentration or the time interval between the first and second peaks to estimate the distance. These schemes require less time with respect to two-way transmission protocols, but they are derived for a 1-D diffusion channel, and the estimation accuracy is not improved significantly. In \cite{noel2014bounds}, a lower bound for distance estimation accuracy is derived for the diffusion channel without any flow and initial drift. In another study, two distance estimation schemes, where emission time is considered as a parameter, are proposed for a 3-D scenario in a diffusion channel  \cite{wang2015distance}. The first scheme uses the peak time of the received signal, and the second scheme, which gives more accurate results, uses the received energy to estimate the distance. The same authors propose an algorithmic distance estimation scheme and two parameter optimization methods in \cite{wang2015algorithmic} for a 3-D diffusion channel. When compared with \cite{wang2015distance}, this scheme has a worse distance estimation performance, requires a more complex receiver and needs more time for estimation. In \cite{lin2019high}, a distance estimation protocol is proposed  to estimate the distance in a 3-D diffusion channel in the presence of additive noise and inter-symbol interference by using maximum likelihood estimation. This proposed method has a high accuracy with a cost of high complexity.

All of the distance estimation methods in the literature consider an ideal microscale channel model where the transmitted molecules do not have an initial velocity, the molecules move according to  Brownian motion, there is no gravity, and the TX transmits and the RX receives the signals perfectly. However, this channel model is not realistic for macroscale MC. Furthermore, there is no distance estimation method for macroscale situations in the literature. In this paper, we propose five distance estimation methods for a practical macroscale scenario, where molecules propagate indoors with an initial velocity and without any constant flow in a 3-D medium, i.e., air. Three of the proposed methods are novel data analysis based methods, and two of them include supervised machine learning (ML) methods. To collect data for the purpose of using them in these methods, an experimental setup similar to the tabletop platform in \cite{farsad2013tabletop} is employed. In this setup, an electric sprayer without a fan is the TX that emits alcohol molecules and a metal-oxide alcohol sensor is the RX. The TX and RX are controlled with a microcontroller via a computer to record the received signals. A novel algorithm is proposed to extract features from the measured signals by the RX. For the first time, the distance estimation is made with supervised ML methods for a practical macroscale MC system. Multivariate linear regression (MLR) and neural network regression (NNR) are used as ML methods which use the extracted features as inputs. The experimental data are used as the training and test data of these methods. Afterwards, the collected data are analyzed and three distance estimation methods are proposed which are less complex but less accurate than ML methods. The first data analysis based method is power based distance estimation, which employs the relation between the distance and the received power to transmitted power ratio. In the second data analysis based method, which is the peak time based distance estimation, the relation between the peak time of the received signal and the distance is exploited. At long distances (170 cm and longer), this method has the best results among the data analysis based methods. Combined distance estimation is the third method, which uses a combination of the power based and peak time based estimation. Furthermore, all of these applied methods are compared by discussing the numerical results. It is shown that ML methods perform better than the data analysis based methods.

The existing theoretical models which are based on the diffusion of the molecules are insufficient to explain the experimental results. It is difficult to consider all of the effects such as turbulent flows and boundary conditions of the molecule movements for deriving an analytical expression to estimate the distance. This gives us the motivation to employ ML and data analysis based methods to estimate the distance as detailed in Section \ref{Motivation}.  Moreover, an analysis is made on the experimental data. Experimental findings show that the movement of the molecules in a practical scenario is also affected by some flows accompanying diffusing molecules. In particular, the analyses reveal that unsteady flows can affect the propagation of the molecules in addition to the Brownian motion, even if there is no fan behind the TX. Correspondingly, a possible trajectory of the transmitted molecules in the communication channel is given. In addition, our analysis based on experimental findings indicate that the  non-linear characteristic of the sensor can cause measurement errors at lower distances. The main result of these analyses is that a fluid dynamics perspective is needed for channel modeling and parameter estimation in MC for macroscale practical scenarios. Also, open research directions are indicated by analyzing the numerical results.

The rest of the paper is organized as follows. In Section \ref{Motivation}, the motivation for using the ML and data analysis based methods is given. In Section \ref{Experimental_Setup}, the experimental setup, which is used for data collection, is explained. The proposed feature extraction algorithm is given in Section \ref{Feature_Extraction}. The supervised ML methods, which are applied to estimate the distance in this paper, are briefly introduced in Section \ref{ML}. In Section \ref{DEM}, the proposed distance estimation methods based on data analysis are given. In Section \ref{Results_Comparison}, the numerical results for the implemented methods are presented and compared. Moreover, received signals and the motion of the molecules are analyzed with a fluid dynamics perspective. Finally, the concluding remarks are given in Section \ref{Conclusion}.


\section{Motivation}
\label{Motivation}
In this section, the motivation of using ML and data analysis based methods for distance estimation in a practical macroscale scenario is given. When the theoretical model of the molecule concentration with respect to time and space is known, the distance between the TX and RX can easily be estimated by measuring other parameters for a MC channel such as molecule concentration, number of transmitted molecules, propagation time of the molecules and the diffusion coefficient. Our experimental setup, which is detailed in Section \ref{Experimental_Setup}, includes a sprayer as the TX, ethyl alcohol (ethanol) as messenger molecules and an alcohol sensor as the RX. There is not a fan behind the TX to create a constant flow and the molecules propagate with an initial velocity in the air. The theoretical models in the literature consider that the transmitted molecules only diffuse in the medium with or without a constant flow. However, these models are not realistic for practical scenarios due to the following reasons. Firstly, boundary conditions and the types of the boundaries, i.e., absorber or reflector, should be taken into account. Secondly, alcohol is initially sprayed with an initial velocity as liquid droplets rather than alcohol molecules. Therefore, only Brownian motion cannot be sufficient to characterize the movement of molecules.

\begin{figure}[H]
	\centering
	\includegraphics[width=0.3\textwidth]{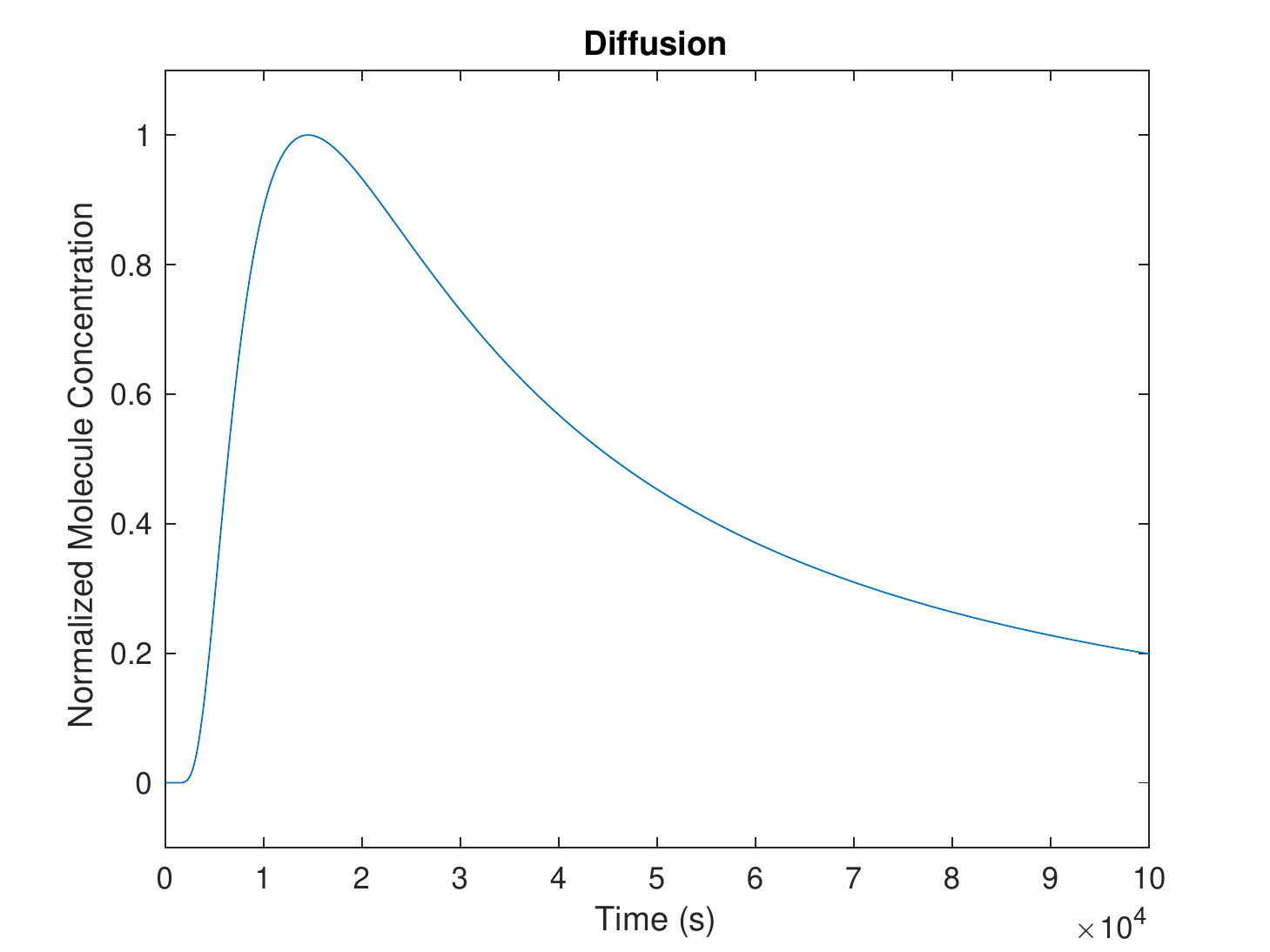}    
	\includegraphics[width=0.3\textwidth]{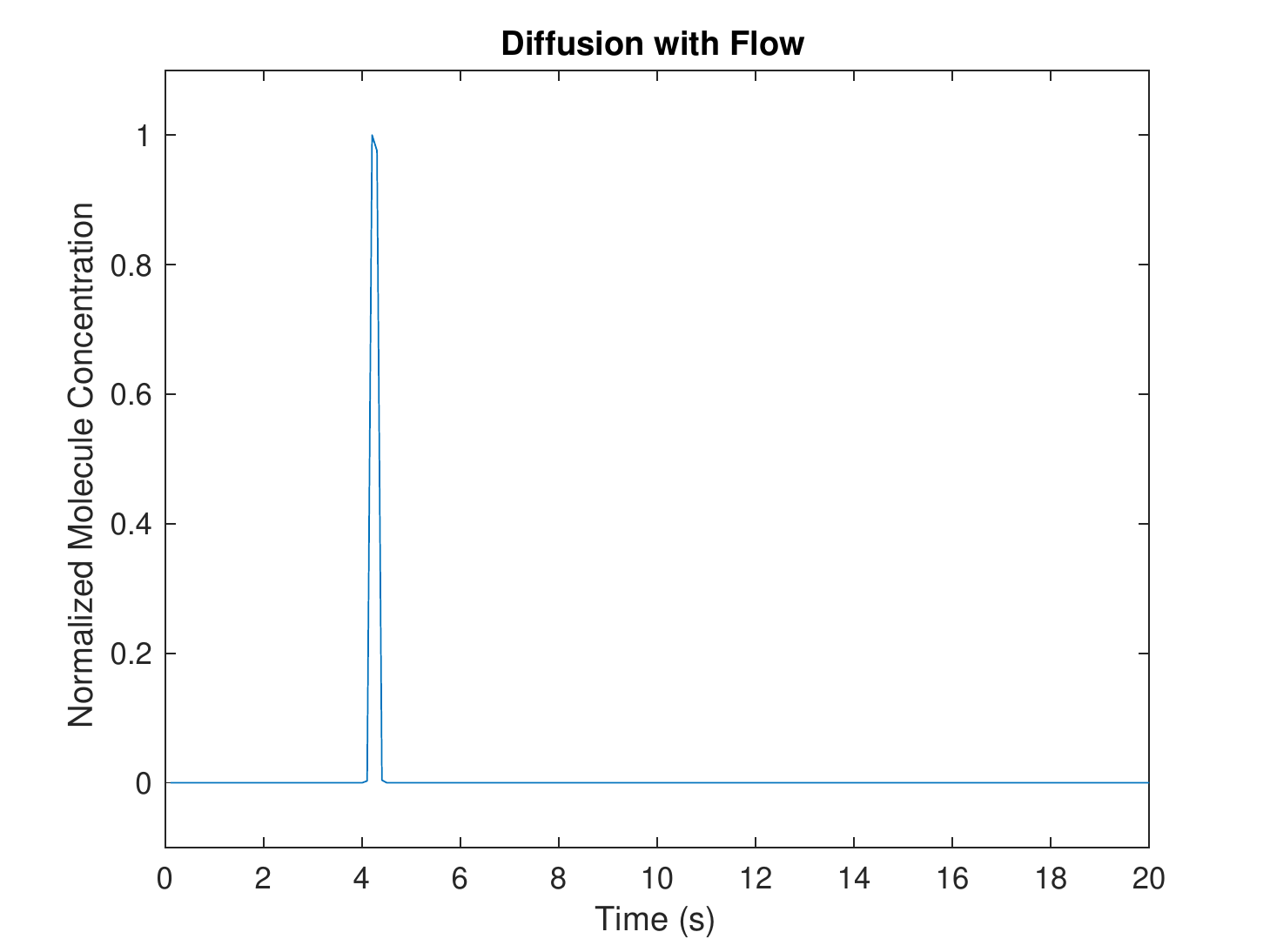}  
	\includegraphics[width=0.3\textwidth]{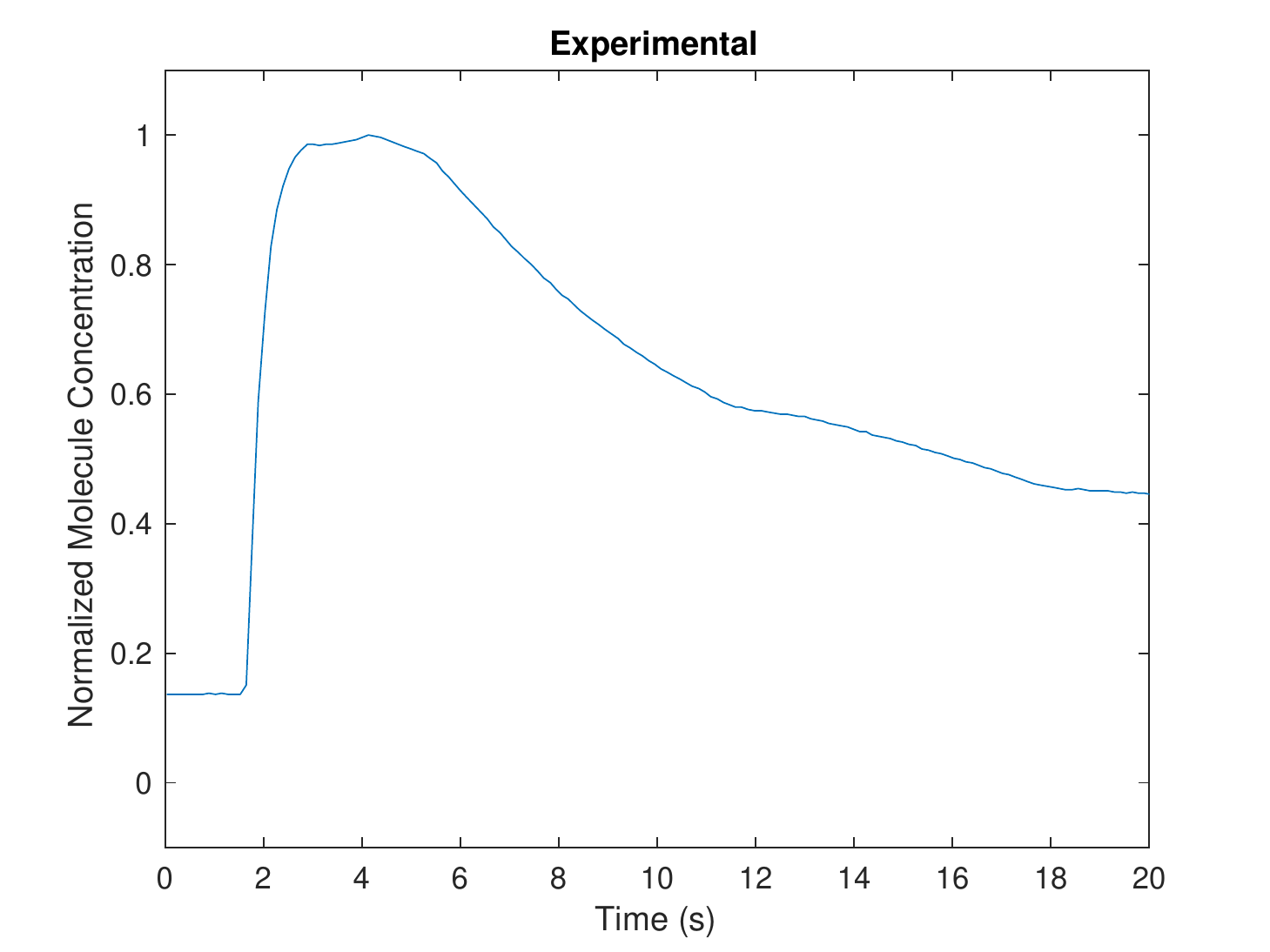} \\
	\scriptsize \hspace{0.6 in} (a) \hspace{1.8 in}  (b) \hspace{1.8 in} (c) \hspace{0.5 in}
	\caption{Comparison of the theoretical models for normalized molecule concentrations as a function of time at $d = 100$ cm with experimental results: (a) Diffusion, (b) Diffusion with flow, (c) Experimental results ($D = 0.1149980886$ $cm^2/s$, $v = 23.53 $ $cm/s$). }
	\label{Motivation_Comp}
\end{figure}

The molecule concentration with respect to time at the RX side does not comply with the existing theoretical models, when the molecules are transmitted with an initial velocity as in our case. The equations for these models are given below in (\ref{Diff}) and (\ref{Diff_flow}) for only diffusion \cite{bossert1963analysis} and diffusion with constant flow  \cite{farsad2014channel}, respectively. 
\begin{equation}
C(t) = \frac{Q}{(4\pi Dt)^{3/2}} e^{-\frac{d^2}{4Dt}}.
\label{Diff}
\end{equation}
\begin{equation}
C(t) = \frac{Qd}{\sqrt{4\pi Dt^3}} e^{-\frac{(vt-d)^2}{4Dt}}.
\label{Diff_flow}
\end{equation}
In these equations, $C(t)$ is the molecule concentration, $ Q $ is the number of the transmitted molecules, $ d $ is the distance from the TX, $ D $ is the diffusion coefficient, $ t $ is the propagation time of the molecules and $ v $ is the velocity of the constant flow in the medium. In Fig. \ref{Motivation_Comp}, a comparison of the theoretical models with the experimental results are given by showing the normalized molecule concentration change in time at a distance $d=100$ cm. In order to see the waveforms more clearly, the normalization is made by dividing $C(t)$ to the peak concentration value. In Fig. \ref{Motivation_Comp}(a), the molecules only diffuse according to (\ref{Diff}). In Fig. \ref{Motivation_Comp}(b), the model in (\ref{Diff_flow}) is applied in a medium with a constant flow, whose velocity is taken as the experimental average velocity  of the molecules at $d=100$ cm.

As shown with the experimental results in Fig.\ref{Motivation_Comp}(c), the models in (\ref{Diff}) and (\ref{Diff_flow}) do not give accurate results for the molecule concentration as a function of time in a practical situation. The diffusion model in (\ref{Diff}) approximates the shape of $C(t)$, but the peak time differs very much from the experimental result due to the effect of the initial velocity. Also, the diffusion with flow model in (\ref{Diff_flow}) gives the peak time approximately, when the average velocity of the molecules is known. However, the shape of the function $C(t)$ is very different from the experimental results. In addition, the solution of the general advection-diffusion equation can be employed to model the molecule concentration 
as given by \cite{mcguiness2018parameter}
\begin{equation}
C(x,y,z,t) = \frac{m_t}{(4\pi t)^{3/2} \sqrt{D_x D_y D_z}} e^{ -\frac{\left(x-(x_0 + v_x t)\right)^2}{4D_x t} -\frac{\left(y-(y_0 + v_y t)\right)^2}{4D_y t} -\frac{\left(z-(z_0 + v_z t)\right)^2}{4D_z t}},
\label{Advection}
\end{equation}
where $m_t$ is the transmitted mass of the molecules, $v_x$, $v_y$ and $v_z$ show the flow velocities at the corresponding $x$, $y$ and $z$ coordinates, respectively, $D_x$, $D_y$, $D_z$ are the diffusion coefficients at the corresponding coordinate given at their subscript and $x_0$, $y_0$ and $z_0$ are the transmission points in three dimensions. In (\ref{Advection}), the estimation of the localized velocity terms with respect to time and space is a difficult problem  which is beyond the scope of this paper. The practical conditions make this estimation problem more complicated due to the reasons given as follows.

In a practical indoor scenario, it is not sufficient to use only Brownian motion due to the effects such as  boundary conditions and the initial velocity of the molecules. It is also very difficult to determine whether the indoor boundaries with various geometries are absorber or reflector. The models in (\ref{Diff}-\ref{Advection}) need to be modified to consider these effects. Furthermore, these modifications may not be sufficient to model the practical indoor scenario, since alcohol molecules are sprayed as liquid droplets and they are subject to gravity and the observed phenomena about the unsteady flows, e.g., turbulent flows, in which the velocity of the droplets is not constant according to time. The details about these flows are given in Section \ref{Results_Comparison}. It is highly difficult to derive an analytical expression, which can take all of the aforementioned effects into account, for channel parameter estimation such as distance. Therefore, it is reasonable to use the alternative practical methods such as data analysis, linear regression and neural networks for estimating the distance accurately. With these methods, these effects on the collected signals are implicitly modeled and employed for distance estimation. Next, the experimental procedure is explained to collect data and implement distance estimation methods.

\section{Experimental Setup and Data Collection}
\label{Experimental_Setup}
\begin{figure}[!t]
	\centering
	\scalebox{0.65}{\includegraphics{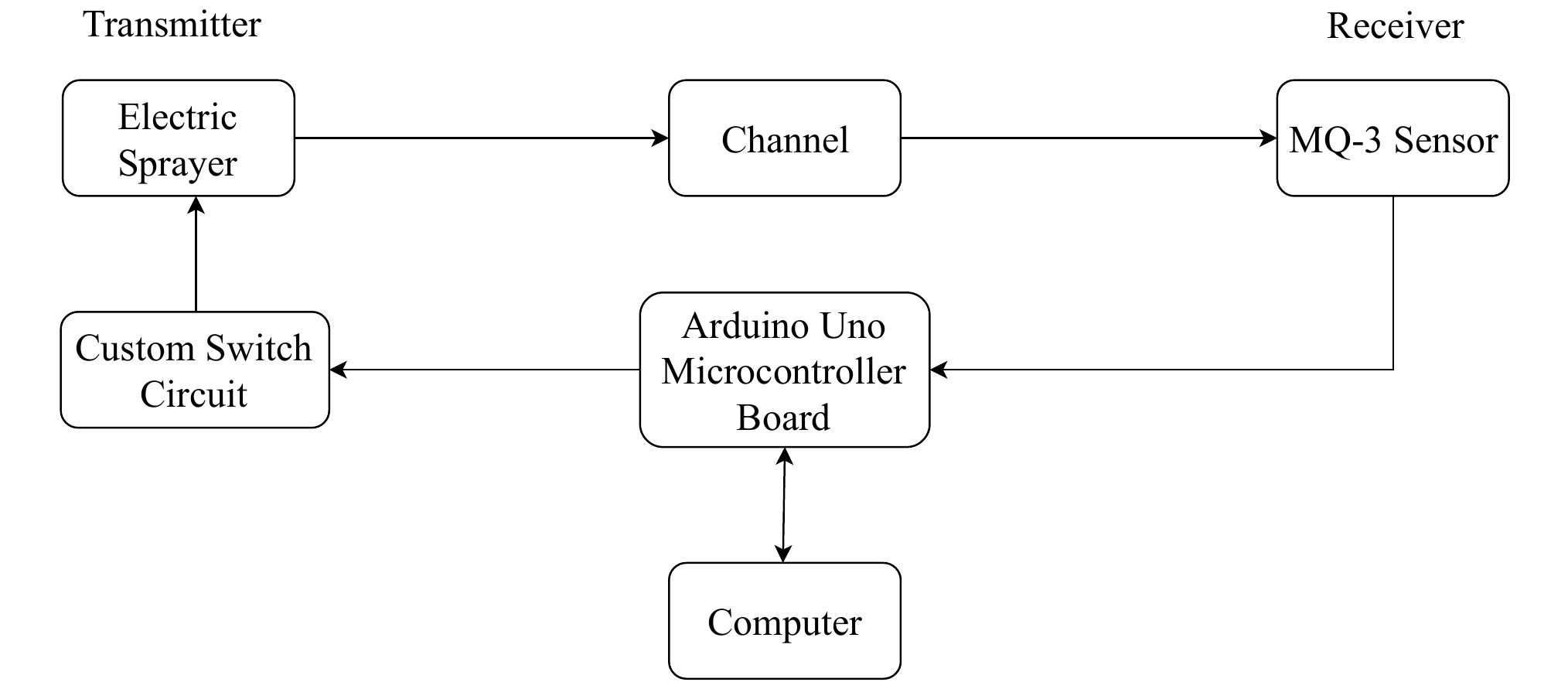}}
	\caption{Block diagram of the experimental setup.}
	\label{TX_RX}
\end{figure}

In this section, the experimental setup of a practical macroscale MC system is explained in order to collect data for distance estimation. This system consists of a transmitter (TX), a receiver (RX) and a propagation channel similar to the tabletop molecular communication system in \cite{farsad2013tabletop}. However, unlike \cite{farsad2013tabletop}, there is no fan in our system. The TX transmits molecular signals by spraying ethyl alcohol with an Instapark electric sprayer. This sprayer has its own battery in it and can be controlled with a microcontroller via a custom switch circuit. The RX receives the molecular signals with an MQ-3  alcohol sensor which gives the best performance among the low-cost metal oxide alcohol sensors \cite{farsad2013tabletop}. The TX and RX are both controlled with an Arduino Uno microcontroller board which is connected to a computer. The block diagram of the system is given in Fig. \ref{TX_RX}. As shown in this diagram, only one microcontroller board and one computer is used for simplicity and perfect synchronization between the TX and RX.

The TX is controlled with a computer via an Arduino microcontroller board so that the spraying duration, i.e., emission time ($T_e$), can be arranged. The TX and RX were $ 100 $ cm high from the ground and the experiments were carried out in a laboratory. After the TX transmits a signal, the molecules begin to diffuse in the air with an initial velocity supplied by the TX. In addition, it is observed that some of the liquid droplets fall to the ground as soon as they are emitted during the experiments. The droplets coming out of a sprayer can be affected by the gravity due to their size \cite{de2017investigation}.  Since the molecules are transmitted as sufficiently large liquid droplets, it is deduced that they are affected by the gravity.

\begin{figure}[htb]
	\centering
	\scalebox{0.15}{\includegraphics{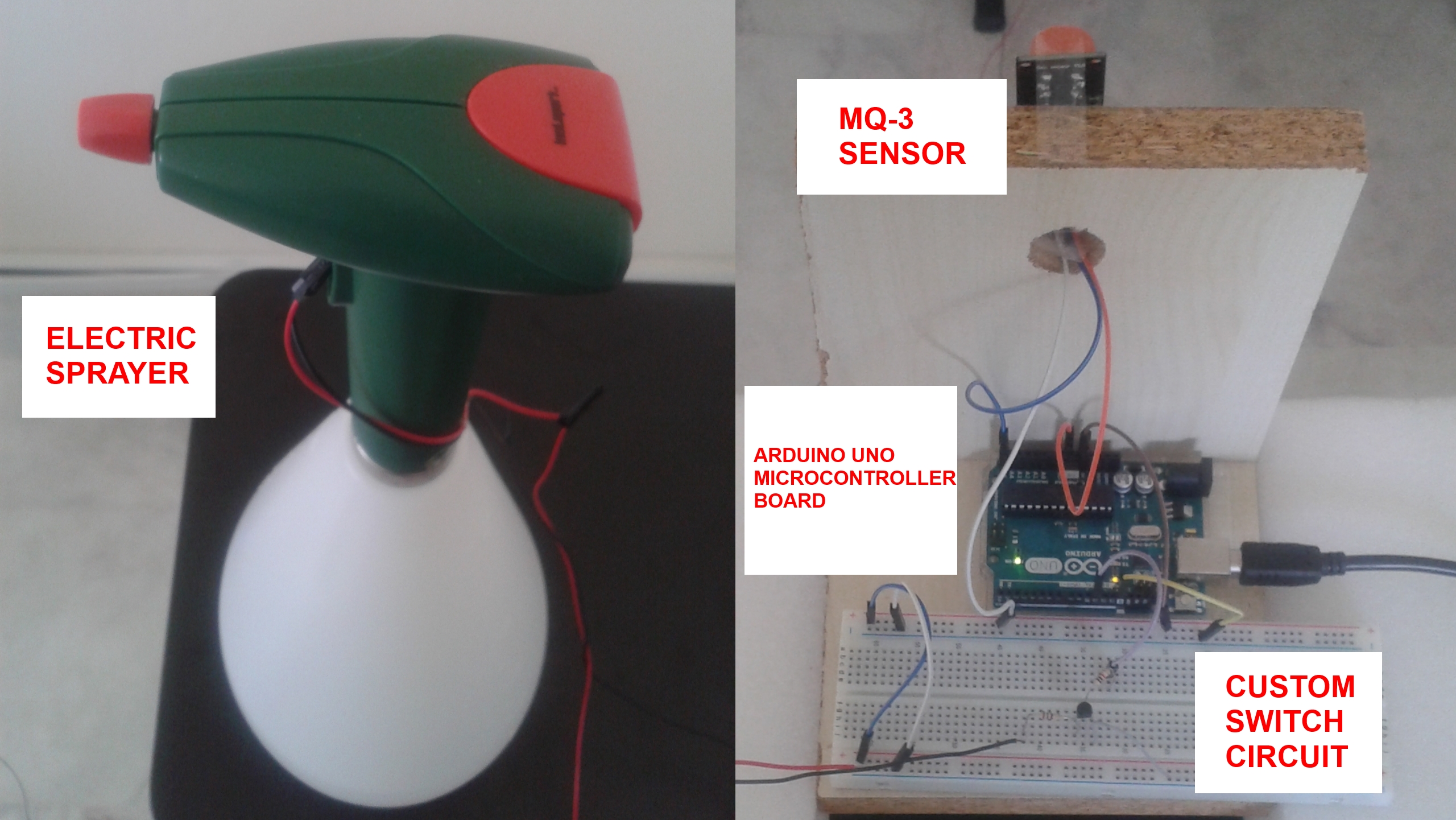}}
	\caption{Components of the transmitter (left) and the receiver (right).}
	\label{Comp}
\end{figure}

When the TX stops transmitting, the RX starts to record the molecular signal for $ 100 $ seconds for distances up to $ 180 $ cm and $ 300 $ seconds for longer distances. The molecule concentration in the environment is sensed by the MQ-3 sensor and analog voltage values corresponding to the measured molecule concentration are sent to the Arduino microcontroller board. This analog signal is digitized with a 10-bit analog to digital converter by the microcontroller board. Finally, the received signal is sent to the computer from Arduino microcontroller board to be recorded. The components of the TX and RX can be seen in Fig. \ref{Comp}. Moreover, the variables which are used in Sections \ref{Experimental_Setup}, \ref{Feature_Extraction} and \ref{DEM} are summarized in Table \ref{Variables}.
\begin{table}[h]
	\caption{Summary of the variables used in Sections \ref{Experimental_Setup}, \ref{Feature_Extraction} and \ref{DEM}.}
	\vspace{0.2cm}
	\centering
	\begin{tabular}{p{45pt}|p{305pt}}
		\hline
		\textbf{Variable}	& \textbf{Definition} \\
		\hline
		$T_e$ & Emission time   \\
		$d$ & Actual Distance \\
		$\hat{d}$ & Estimated distance\\
		$ A_{o} $ & Initial offset level \\
		$ A_{thr} $ &  Threshold amplitude \\
		$ K $ & Detection threshold \\
		$ C[n] $ & Measured molecule concentration \\
		$y[n]$ & Output of the moving average filter\\
		$W_1$ & The number of samples before the $n^{th}$ sample of the moving average filter\\
		$W_2$ & The number of samples after the $n^{th}$ sample of the moving average filter\\

		$t_{peak}$ & Peak time\\
		$C_{peak}$ & Peak molecule concentration\\
		$t_{low}$ & Time showing the 10\% reference point on the rising edge of the measured signal \\
		$C_{low}$ & Molecule concentration showing the 10\% reference point on the rising edge of the measured signal\\
		$t_{high}$ & Time showing the 90\% reference point on the rising edge of the measured signal \\
		$C_{high}$ & Molecule concentration showing the 90\% reference point on the rising edge of the measured signal\\
		$R$ & Rise time on the rising edge of the measured signal \\
		$\Delta C$ & Molecule concentration level difference between $C_{high}$ and $C_{low}$ \\
		$G$ & Gradient on the rising edge of the measured signal \\
		$N_R$ & Number of the received signal samples up to $t_{peak}$\\
		$E_R$ & Received energy up to $t_{peak}$\\
		$P_R$ & Received power 	up to $t_{peak}$\\
		$P_T$ & Transmitted power\\
		$\overline{P_R} $ & Average received power\\ 	
		$\overline{P_T} $ & Average transmitted power\\
		$a_1$, $b_1$ & Curve fitting parameters of power based estimation\\
		$\overline{t_{peak}} $ & Average peak time\\
		$a_2$, $b_2$ & Curve fitting parameters of peak time based estimation\\		
		\hline
	\end{tabular}
	\label{Variables}
\end{table}
During the experiments, only one single puff is transmitted from the TX to estimate the distance for one measurement. The transmissions are repeated for three different emission times as $T_e = \{0.25, 0.5, 0.75\}$ s and eleven different distances between $ 100 $ cm and $ 200 $ cm with $ 10 $ cm steps. The parameter ranges and replication numbers for the experiment are given in Table \ref{Parameters}. A break was taken for at least 5 minutes between consecutive measurements. During this break, the laboratory is ventilated by opening the door and the window to reduce the molecule concentration level inside the laboratory. By the help of this ``break and ventilate" procedure, the effect of the remaining molecules from previous transmissions is minimized. Without applying this procedure, the noise level increases and thus, it gets more difficult to detect the signal by the RX side. In order to estimate the distance between the TX and RX, it is essential to extract the features from the signals to employ them as inputs to ML methods, as discussed in the next section.
\begin{table}[h]
	\caption{The experimental parameters and their ranges.}
	\vspace{0.2cm}
	\centering
	\begin{tabular}{cc}
		\hline
		\textbf{Parameter}	& \textbf{Value} \\
		\hline
		$T_e$ & \{0.25, 0.5, 0.75\} s  \\
		$d$ & \{100, 110, 120,..., 190, 200\} cm \\
		Replication for each $T_e$ in every distance & 10 \\
		Replication for each distance & 30 \\
		Number of total measurements & 330 \\                                             
		\hline
	\end{tabular}
	\label{Parameters}
\end{table}
\section{Feature Extraction}
\label{Feature_Extraction}
\begin{figure}[!htb]
	\centering
	\includegraphics[width=0.3\textwidth]{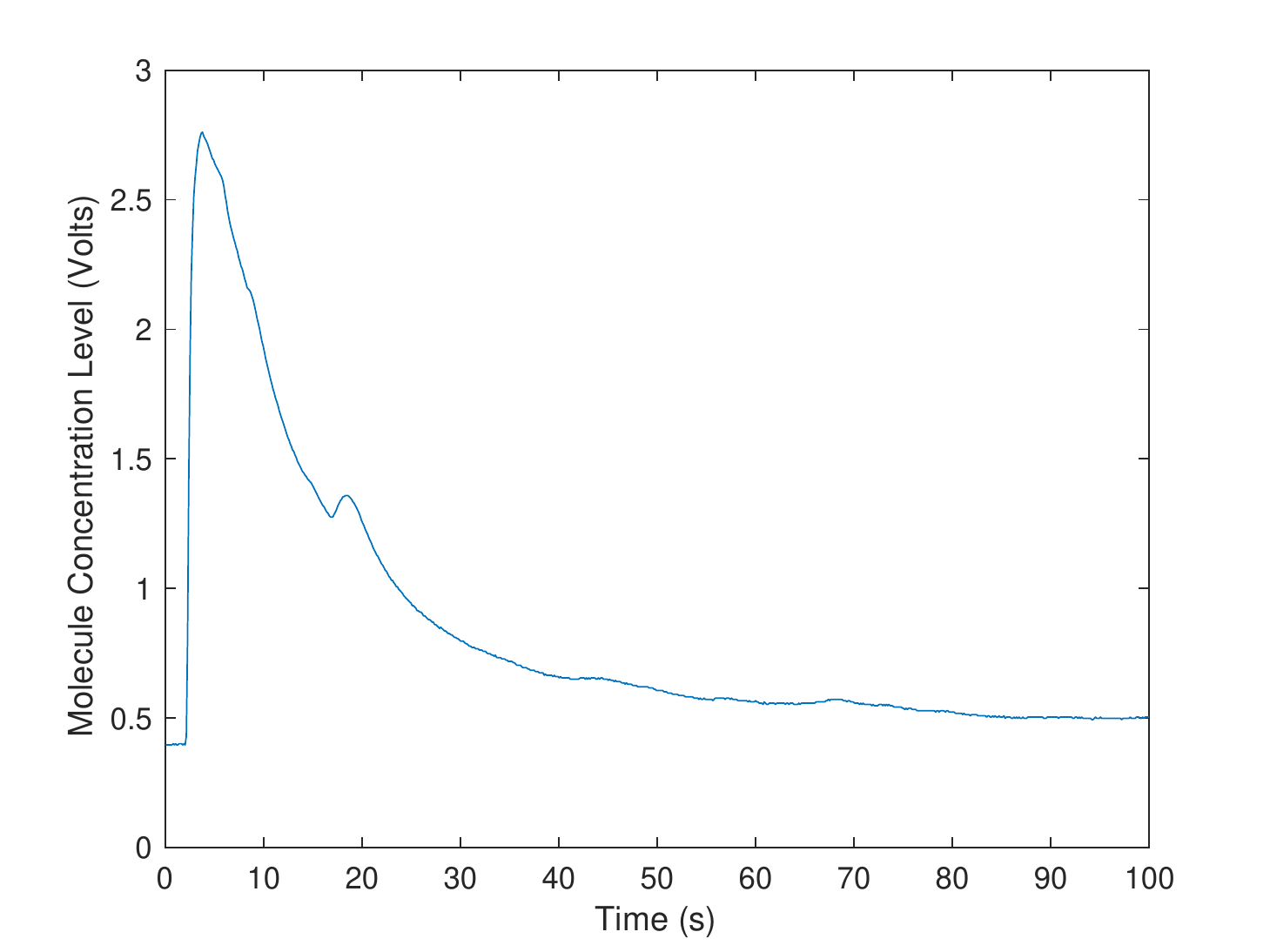}    
	\includegraphics[width=0.3\textwidth]{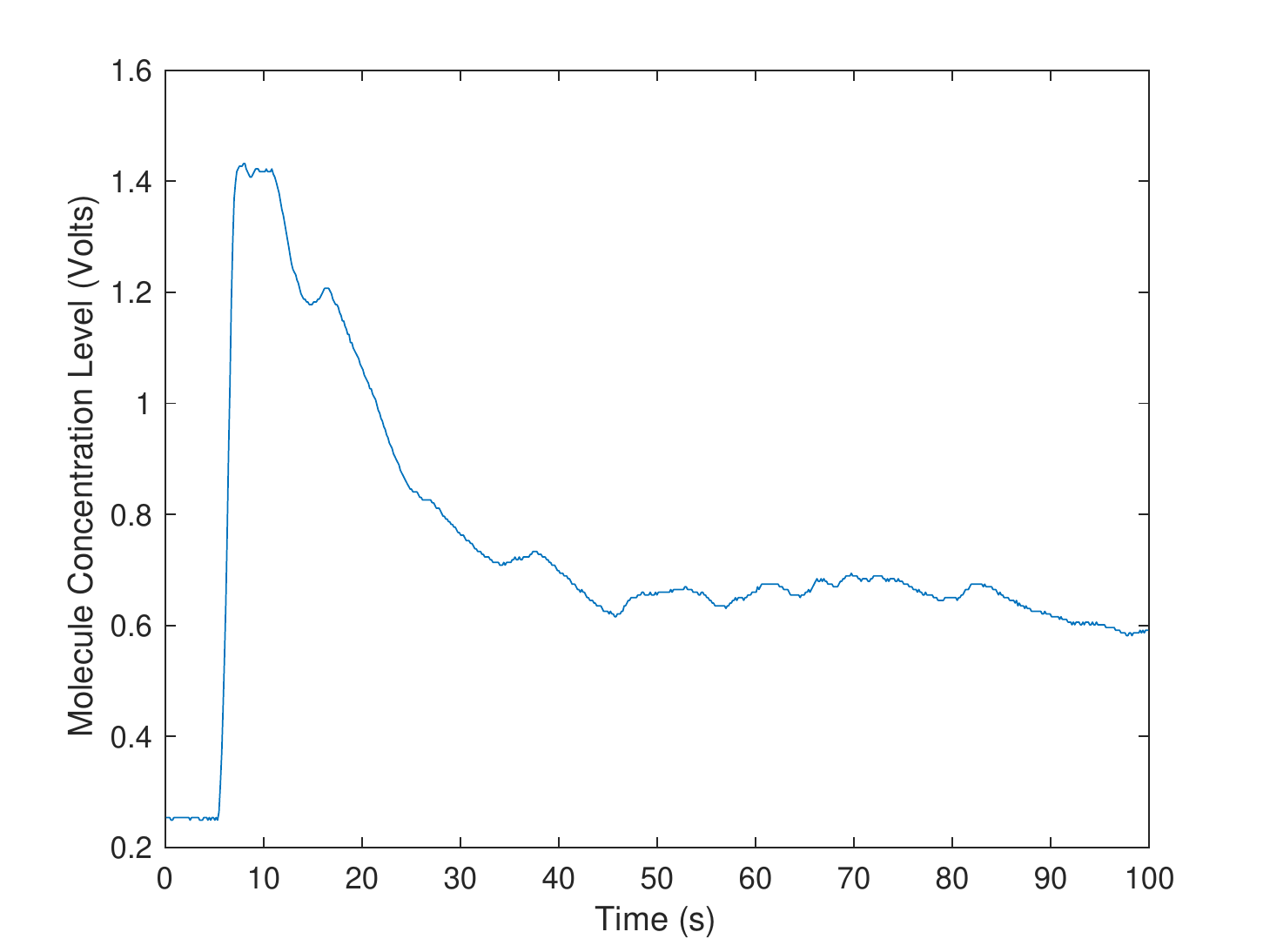}  
	\includegraphics[width=0.3\textwidth]{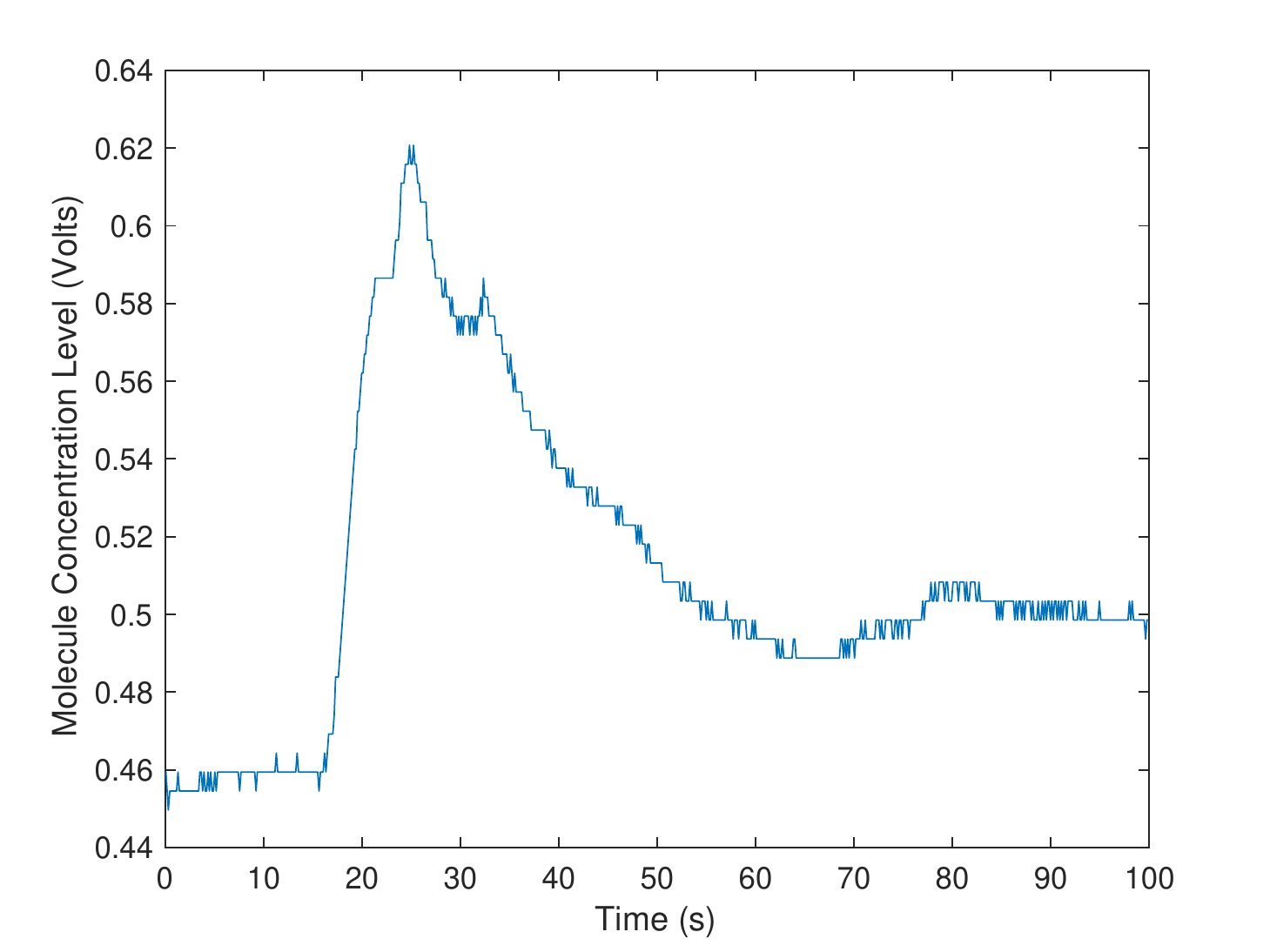} \\
	\scriptsize \hspace{0.6 in} (a) \hspace{1.8 in}  (b) \hspace{1.8 in} (c) \hspace{0.5 in}
	\caption{Received signal at (a) $d = 100$ cm , (b) $d = 160$ cm (c) $d = 200$ cm.}
	\label{d_features}
\end{figure}

In this section, a novel feature extraction algorithm is proposed. This algorithm processes the received molecular signals to produce features, which can be defined as the input variables characterizing the output variable to be estimated for ML algorithms \cite{christopher2016pattern}. The choice of the features is important, since it effects the performance of the ML algorithms. Eight features which are $ t_{low}$, $C_{low}$, $R$, $\Delta C$, $G$, $t_{peak}$, $C_{peak}$ and $E_R$ are extracted from the collected data. Three of the measured signals at $100$, $160$ and $200$ cm are shown in Fig. \ref{d_features} in order to justify the choice of the selected features. In this figure, the delay time, which is defined as the beginning time of the detection of the transmitted molecules, increases, as the distance increases. $t_{low}$ gives a more reliable information about the delay time due to the fluctuations of the signal. $C_{low}$ also decreases, as the distance increases. Since the signal attenuates as the distance increases, the peak points ($t_{peak}$, $C_{peak}$) and the received energy ($E_R$) of the signals contain information about the distance. Furthermore, as the distance increases, the slope of the signals' rising edge decreases. This shows that $R$, $\Delta C$ and $G$ are informative features. To the best of our knowledge, the proposed feature extraction algorithm which is summarized in Fig. \ref{Features}, is the first one that processes the real measured molecular signals.

\tikzstyle{startstop} = [rectangle, rounded corners, minimum width=3cm, minimum height=1cm,text centered, draw=black, fill=red!30]
\tikzstyle{io} = [trapezium, trapezium left angle=70, trapezium right angle=110, minimum width=1cm, minimum height=1cm, text centered, text width=3.5cm, draw=black, fill=orange!30]
\tikzstyle{decision} = [diamond, draw, fill=green!20, 
text width=4.5em, text badly centered, node distance=3cm, inner sep=0pt]
\tikzstyle{block} = [rectangle, draw, fill=blue!20, 
text width=3.5cm, text centered, rounded corners, minimum height=2cm]
\tikzstyle{line} = [draw, -latex']
\tikzstyle{cloud} = [draw, ellipse,fill=red!20, node distance=3cm,
minimum height=2em, text width=3.5cm, text centered]

\begin{figure}[!t]
	\centering
	\scalebox{0.6}{
		\begin{tikzpicture}[node distance = 3cm, auto]
		\node [startstop] (0) {Start};
		\node [io, right of=0, xshift=2.5cm] (1) {Record the received signal from the MQ-3 sensor};
		\node [block, right of=1, xshift=2.5cm] (2) {Determine the initial offset level ($ A_{o} $)};
		\node [block, below of=2, yshift=-2.5cm] (3) {Set the threshold, $K$ for detection. $ K = A_{o} + A_{thr}$};
		\node [block, left of=3,  xshift=-2.5cm] (4) {Smooth the signal by a moving average filter};
		\node [decision, below of=4, yshift=-2.5cm] (5) {Detection: Is there any sample above the threshold, $ K $ ?}; 
		\node [block, left of=5, xshift=-2.5cm] (6) {Wait for the new received signal};
		\node [block, right of=5, xshift=2.5cm] (7) {Detect the first peak point of the signal};
		\node [block, below of=7, yshift=-1.5cm] (8) {Detect the nearest local minimum point before the first peak of the signal};
		\node [block, left of=8, xshift=-2.5cm] (9) {Extract the features:  $ t_{low} $, $ C_{low} $, $ R $, $ t_{peak} $, $ C_{peak} $,  $ \Delta C $, $ G$};
		\node [block, left of=9, xshift=-2.5cm] (10) {Calculate the energy of the received  signal ($ E_R $) up to $t_{peak}$ s};
		\node [io, below of=10, yshift=-1.5cm] (11) {Send the values of the features to machine learning methods as inputs};
		\node [startstop, right of=11, xshift=2.5cm] (12) {Stop};
		\path [line] (0) -- (1);
		\path [line] (1) -- (2);
		\path [line] (2) -- (3);
		\path [line] (3) -- (4);
		\path [line] (4) -- (5);
		\path [line] (5) -- node [near start] {no} (6);
		\path [line] (6) -- (1);	
		\path [line] (5)-- node [near start] {yes} (7);	
		\path [line] (7) -- (8);	
		\path [line] (8) -- (9);
		\path [line] (9) -- (10);		
		\path [line] (10) -- (11);	
		\path [line] (11) -- (12);
		\end{tikzpicture}
	}
	\caption{Feature Extraction Algorithm.}	
	\label{Features}
\end{figure}
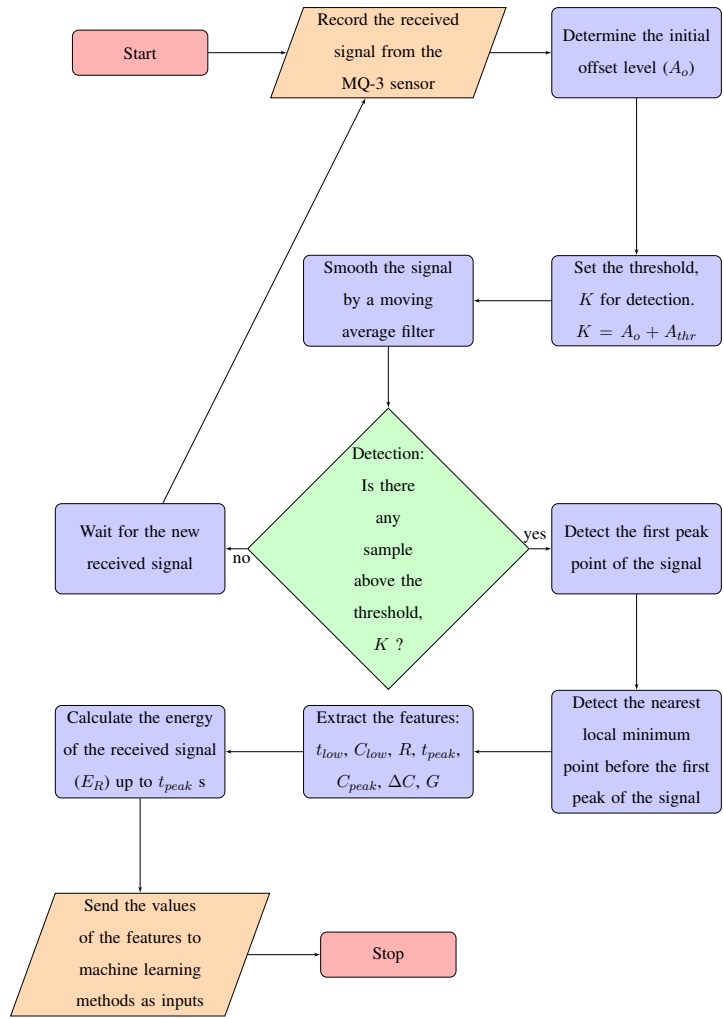

\begin{figure}[!t]
	\centering
	\scalebox{0.6}{\includegraphics{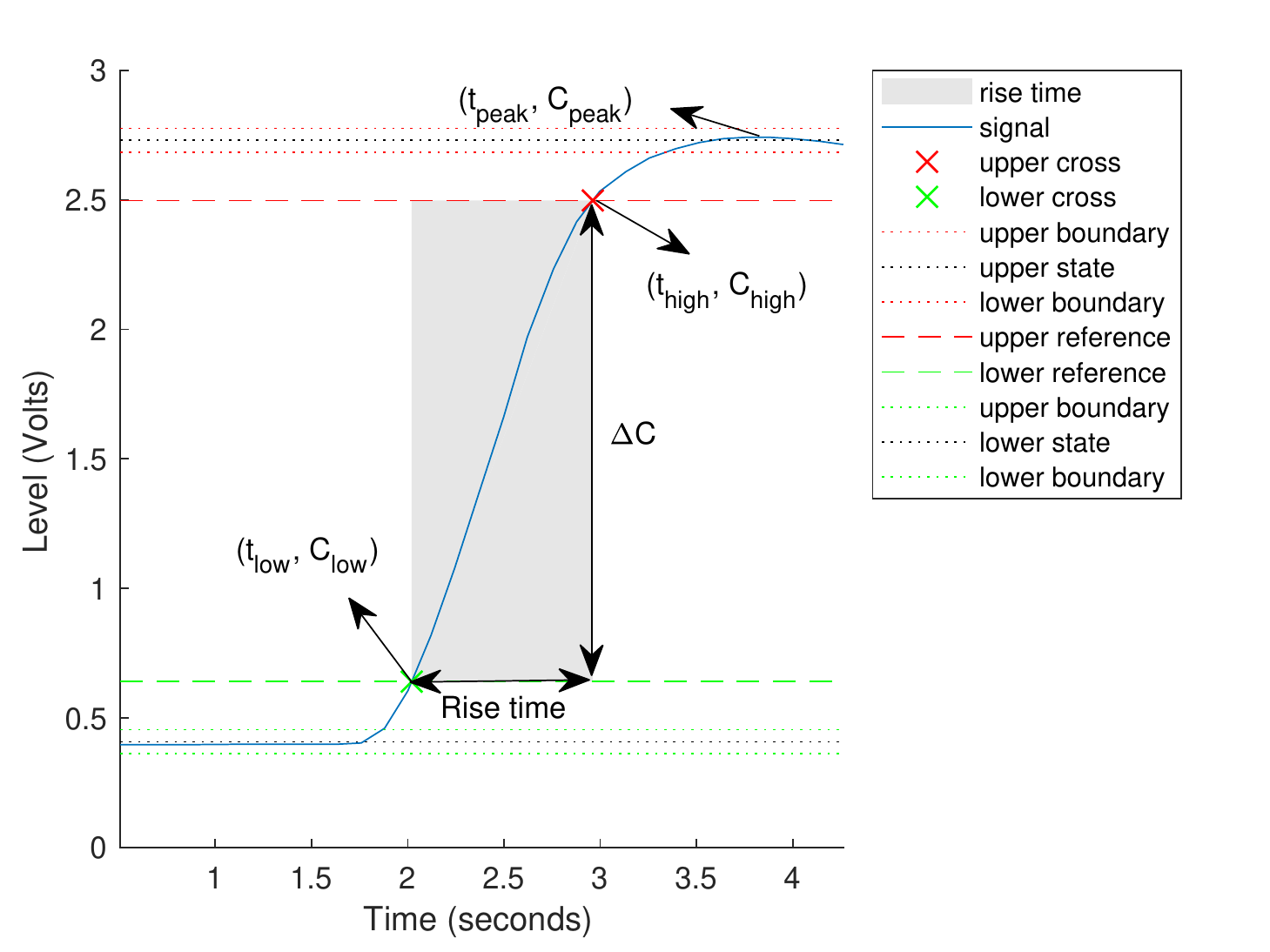}}
	\caption{Extracted features from the received signal.}
	\label{Feat_plot}
\end{figure}

As the first step of the feature extraction algorithm, $ A_{o} $, which is defined as the initial offset level characterized by the average of the measured sensor voltage, is determined. $ A_{o} $ is calculated by averaging the first $p$ samples of the signal, before the transmitted molecules start to be sensed by the RX. In order to avoid considering the transmitted molecules by the TX, $ p $ is selected as $ 5 $ empirically for our experimental scenario. The molecule concentration level is measured as the voltage values received from the MQ-3 sensor. $ A_{o} $ corresponds to the average molecule concentration level which is measured before the transmitted signal arrives. After $ A_{o} $ is determined, the detection threshold is given by the equation
\begin{equation}
K = A_{o} + A_{thr},
\end{equation}
where $ A_{thr} $ is set to 0.1 volts. The value of $ A_{thr} $ is chosen sufficiently high according to our observations in order to detect the transmitted signal. The derivation of a selection criterion for $A_{thr}$ requires to handle the related detection theory together with the probability distributions of the observed molecular signals. This is beyond the scope of this paper. Subsequently, smoothing the received signal is needed due to the fluctuations in the signal orginated from the random movements of the molecules. The signal is smoothed by filtering with a moving average filter whose input-output  relationship is given by \cite{oppenheim1999discrete}
\begin{equation}
y[n] = \dfrac{1}{1+W_1+W_2} \sum_{k = -W_1}^{W_2} C[n-k].
\end{equation}
Here, ($W_1 + W_2 + 1$) gives the window size which is chosen as $7$ by setting $W_1 = W_2 = 3$. This window size is chosen empirically in order to eliminate the ripples on the signal without changing its general shape. Increasing the window size makes the signal smoother. However,  if the the window size is greater than $7$, the values of the extracted features begin to diverge from their original values. Subsequent to smoothing operation, the detection block is accomplished. In this block, a decision is made whether the signal is above the predetermined $K$ or not. If there is no detection, then the RX waits for the new signal. Otherwise, the first peak of the signal is detected. According to our observations, there can be more than one peak in the received signal, after the redundant peaks are eliminated by the smoothing filter. The number of these peaks increase as the distance increases, since the variance of the time distribution that the molecules take to reach the sensor is getting larger. The first peak time is considered as the reference time to calculate the features, since it is assumed that the majority of the molecules arrive at the RX until the first peak time. In order to find the first peak point of the signal, the first derivative of the signal's samples is taken. The first negative sample index at the point where the sign of the first derivative changes from positive to negative values gives the peak of the signal. After that, the nearest local minimum point before the first peak of the signal is detected. The same method to find the peak point of the signal can be used to find the first minimum point of the signal. When the signal is inverted, the first peak point gives the first minimum point of the signal. The features extracted from the first peak of the signal are shown in Fig. \ref{Feat_plot}. Here, the peak time and molecule concentration values are recorded as $t_{peak}$ and $C_{peak}$, respectively. Afterwards, the part of the signal which is between the peak and minimum points is treated as the rising edge of a pulse. The positive-going bilevel waveform consisting of high and low levels is obtained to extract the information which can change according to distance. As illustrated in Fig. \ref{Feat_plot}, $ 10\% $ reference point of the signal gives two features as $t_{low}$ and $C_{low}$ and similarly $ 90\% $ reference point of the signal is chosen to represent $t_{high}$ and $C_{high}$. These reference points are found by calculating the points at $ 10\% $ and $ 90\% $ of the time indices for the time interval between the peak and minimum points. The rise time ($R$) of this waveform is calculated by measuring the time between $t_{low}$ and $t_{high}$.  Another extracted feature for this signal is $\Delta C$ which is defined as the difference in the amplitude level during the rise time measurement points.

The seventh feature is the gradient, which is derived by $ G = \Delta C/R$. $G$ gives the change during the rise time, which can be discriminating according to different distances. Finally, the received energy up to $ t_{peak} $ is calculated by the formula
\begin{equation}
	E_R = \sum_{n = 1}^{N_R} |C[n]-A_{o}|^2.
\end{equation}
In this formula, the received molecule concentration level is decreased by the existing noise level to eliminate its effect. Furthermore, the energy is found up to $t_{peak}$, since the time-dependent function of the concentration values has a long tail and it gets much longer, as the distance increases. In the next section, the ML algorithms that is used in this paper is introduced.

\section{Supervised Machine Learning Methods for Distance Estimation}
\label{ML}
ML has a key role in science and engineering, since it facilitates the estimation of the desired outputs without changing the system model for changing inputs. It provides the ability to improve the estimation models adaptively which is called ``learning". When the inputs and outputs are known during the training of the ML system, it is called ``supervised" learning. In supervised ML, classification and regression are two ways of estimation where the former tries to estimate the discrete results for a discrete output and the latter deals with the estimation of the results of a continuous output. In this paper, ML techniques, such as MLR and NNR are used for the first time to estimate the distance between a TX and an RX in a MC system. Next, the mentioned ML methods are introduced briefly.
\subsection{Multivariate Linear Regression}
MLR is a simple method to estimate the output according to the model equation given below
\begin{equation}
\hat{d} = \theta_0 + \sum_{j = 1}^{m} \theta_j x_j,
\label{linear_reg}
\end{equation}
where $\theta_0$ is the bias, $ m $ is the number of the features, $x_j$'s and $\theta_j$'s represent the features and the unknown coefficients, respectively \cite{friedman2008elements}. (\ref{linear_reg}) can also be expressed in matrix form as
\begin{equation}
\mathbf{\hat{d}} = \mathbf{x} \boldsymbol{\theta} ,
\label{LR_matrix}
\end{equation}
where $\mathbf{x}$ is an $N \times (m+1)$ matrix of features, $ \boldsymbol{\theta} $ is a  column vector of $(m+1)$ elements representing the coefficients, $ \mathbf{\hat{d}} $ is a column vector of $N$ elements showing the estimated distance and  $N$ is the number of samples. Each row of $\mathbf{x}$ shows one sample having an extra $1$ as the first element of each row. $N$ represents the number of training samples for the training period and the number of test samples while the distance is estimated using (\ref{LR_matrix}). The model coefficients are chosen during the training period to minimize the cost function given by the formula below
\begin{equation}
J(\theta) = \frac{1}{N}\sum_{i=1}^{N} (d_i - \hat{d}_i)^2,
\label{cost}
\end{equation}
By using the least squares method in the training period, the coefficients are determined by
\begin{equation}
\boldsymbol{\theta} = ( \mathbf{x}^T  \mathbf{x})^{-1}  \mathbf{x}^T \mathbf{d},
\label{theta}
\end{equation}
where $ \mathbf{d} $ is a column vector with $N$ elements showing the actual distances for training. After finding the coefficients by the training according to (\ref{theta}), the distance can be estimated by (\ref{LR_matrix}) where $N$ shows the number of test samples.

\subsection{Neural Network Regression}
\begin{figure}[!htb]
	\centering
	\scalebox{0.6}{\includegraphics{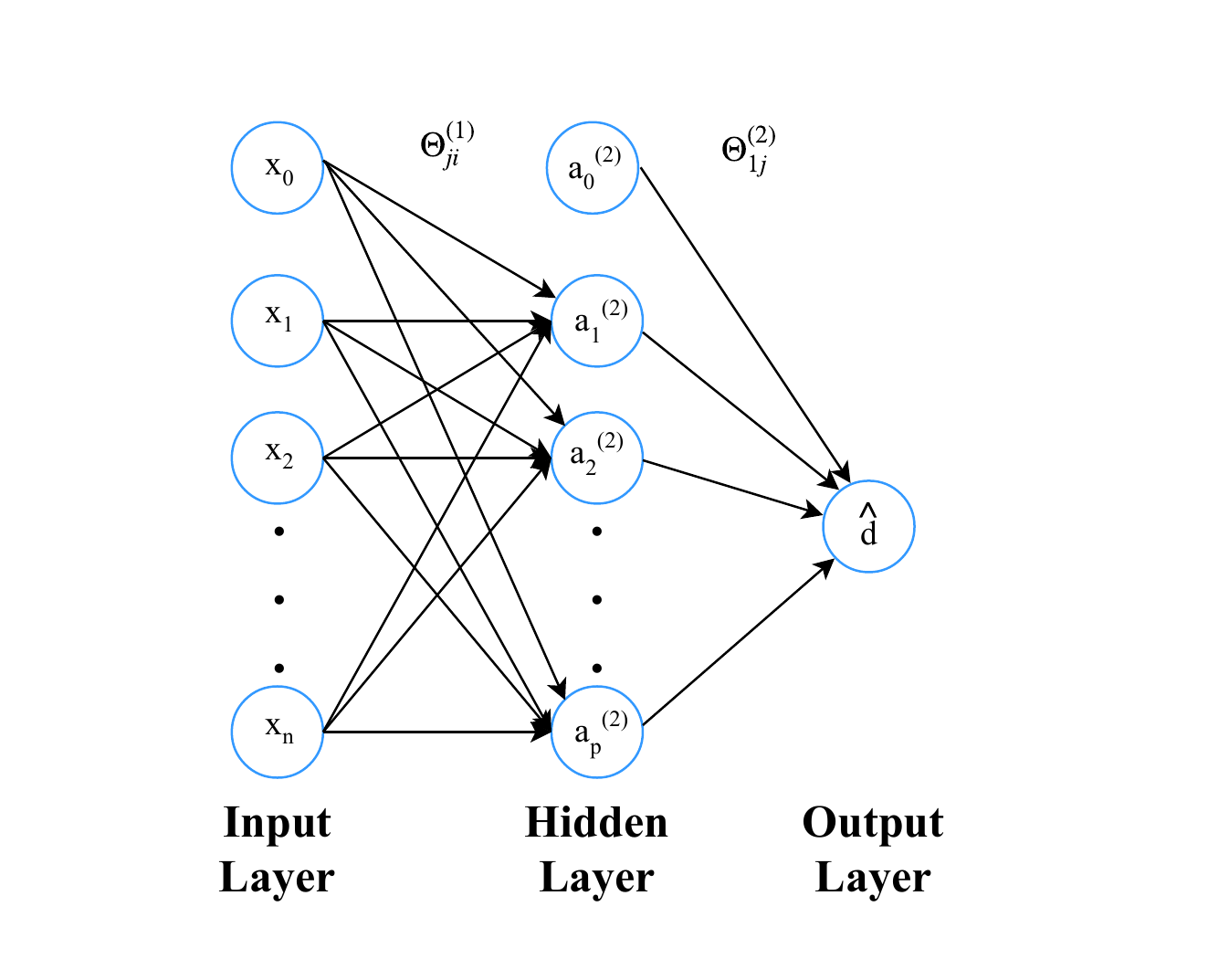}}
	\caption{General structure of a neural network for regression with one output node and a single hidden layer.}
	\label{NN}
\end{figure}

NNR is a ML method inspired by the neurons in the brain. It has a layered structure which includes an input layer, one or more hidden layers and an output layer. These layers consist of nodes like neurons connecting the input layer to the output layer with weights \cite{friedman2008elements}. As illustrated in Fig. \ref{NN}, the weights ($\Theta$'s) are arranged as the elements of a function that maps the input values ($x_i$) to an output value, which is the estimated distance ($\hat{d}$) in our case, with minimum error. 

The relation between the input layer and the hidden layer ($a_j$) can be given by
\begin{equation}
a_j^{(k)} = g\left(\sum_{i=0}^{n} \theta_{ji}^{(k)} x_i\right),
\end{equation}
where $\theta_{ji}^{(k)}$ shows the weight that connects the $i^{th}$ node to the $j^{th}$ node between the layers $(k)$ and $(k+1)$, $a_j^{(k)}$ is the $j^{th}$ hidden node in the $k^{th}$ layer, $x_i$ is the input values defined from $0$ to $n$ and $g(.)$ is the activation function. The relation between the hidden layer and the output layer can be defined as
\begin{equation}
\hat{d} = g\left(\sum_{j=0}^{p} \theta_{ji}^{(k)} x_i\right),
\label{NNR_est}
\end{equation}
where the number of the hidden nodes in the hidden layer is defined from $0$ to $p$. Here, the input and hidden nodes with the subscript ``$0$" are defined as the bias units. In this paper, the hyperbolic tangent sigmoid function is used as the activation function which is given by
\begin{equation}
g(z) = \dfrac{2}{1+e^{-2z}}  - 1.
\end{equation}

The training of the neural network can be made with backpropagation algorithms. In these algorithms, the weights are  initialized generally with random values and the output of the network is calculated in the forward direction. Then, using the output and the values of the hidden nodes, the errors are calculated in the reverse direction and the weights are arranged iteratively to obtain minimum error between the estimated and actual output values. In this paper, Levenberg-Marquardt (LM) backpropagation algorithm, which is one of the most popular and fastest algorithms, is employed to train the neural network \cite{hagan1994training}. Before giving the update mechanism of the LM algorithm, some definitions are made as follows. A pattern is defined as the number of the input-output pairs of the neural network. Each different path from the input layer to the output layer gives a different pattern which is indexed from $ 1 $ to $ r $. The training error for the $ i^{th} $ pattern is given by
\begin{equation}
e_i = r_i - o_i,
\end{equation}
where $ r_i $ is the desired output and the $ o_i $ is the trained output of the neural network for the $ i^{th} $ pattern. Moreover, the error vector ($ \mathbf{e} $) is given as
\begin{equation}
\mathbf{e} = 
\begin{bmatrix}
e_1 & e_2 & \dots & e_r \\
\end{bmatrix}^T,
\end{equation}
where $e_i$ is the training error for the $ i^{th} $ pattern. Similarly, the sum of the squares function is given by
\begin{equation}
E(\boldsymbol{\theta}) = \frac{1}{2} \sum_{i=1}^{r} e_i^2,
\end{equation}
where  $ \boldsymbol{\theta} $ is the weight vector of $N$ elements representing all of the weights in the neural network. The LM algorithm updates these weights to converge the minimum training error by minimizing $ E(\boldsymbol{\theta}) $. For this operation, Jacobian matrix is used which can be given as
\begin{equation}
\mathbf{J} = 
\begin{bmatrix}
\frac{\partial e_1}{\partial \theta_1} & \frac{\partial e_1}{\partial \theta_2} & \dots & \frac{\partial e_1}{\partial \theta_N} \\
\frac{\partial e_2}{\partial \theta_1} & \frac{\partial e_2}{\partial \theta_2} & \dots & \frac{\partial e_2}{\partial \theta_N} \\
\vdots & \vdots & \ddots & \vdots \\
\frac{\partial e_r}{\partial \theta_1} & \frac{\partial e_r}{\partial \theta_2} & \dots & \frac{\partial e_r}{\partial \theta_N} \\
\end{bmatrix},
\end{equation}
where  $ \theta_i $ represents the weights defined from $1$ to $N$. The weight update rule of the LM algorithm in matrix form is given by
\begin{equation}
\boldsymbol{\theta}_{l+1} = \boldsymbol{\theta}_{l} - \left( \mathbf{J}_l^T \mathbf{J}_l + \mu \mathbf{I} \right)^{-1} \mathbf{J}_l \mathbf{e}_l,
\end{equation}
where $l$ is the index of iterations,  $ \mathbf{I} $ is the identity matrix, $\mu$ is a positive coefficient related with the learning rate. 

In LM algorithm,  $ E(\boldsymbol{\theta}) $ is computed and checked to see, if it decreases. If  $ E(\boldsymbol{\theta}) $ decreases, then $\mu$ is reduced by a step size $\gamma$ which is a user defined parameter. Otherwise, $\mu$ is increased by $\gamma$. The algorithm continues to iterate until it reaches to a predetermined threshold value for $ E(\boldsymbol{\theta}) $ \cite{hagan1994training}. After the weights are determined by the training with LM algorithm, the distance can be estimated by using (\ref{NNR_est}). Next, novel distance estimation methods based on data analysis are introduced.

\section{Distance Estimation Methods Based on Data Analysis}
\label{DEM}
The features extracted by the proposed algorithm in Section \ref{Feature_Extraction} are employed to derive novel distance estimation methods in this section. For the first method, the relation among the average received power, average transmitted power and the distance are exploited. Next, the average peak time and distance relation is used for the second estimation method. Finally, these two methods are combined to derive the third distance estimation method.

\subsection{Power Based Distance Estimation}
\label{PBDE}
By using the received energy up to $t_{peak}$, which is calculated in Section \ref{Feature_Extraction}, the received power ($P_R$) is calculated with the formula
\begin{equation}
P_R = \frac{E_R}{N_R} = \frac{1}{N_R}\sum_{n = 1}^{N_R} |C[n]-A_{o}|^2.
\end{equation}
$P_R$ is averaged over $ 10 $ measurements for each $T_e$, i.e., $0.25$ s, $0.5$ s and $ 0.75 $ s, at each distance. Hence, we obtain three average power ($ \overline{P_R} $) values according to three different $T_e$'s for each distance. 

\begin{figure}[!b]
	\centering
	\scalebox{0.5}{\includegraphics{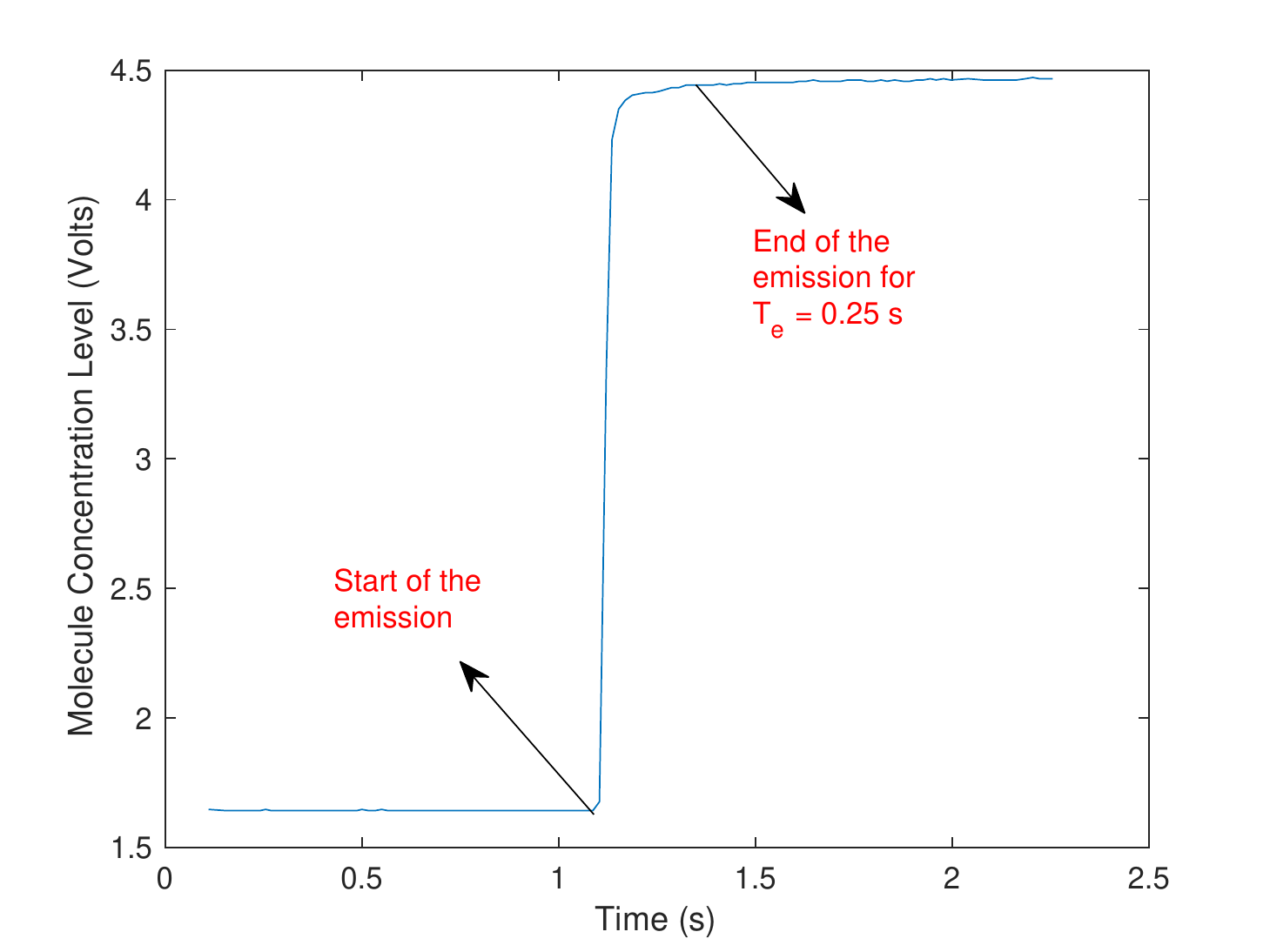}}
	\caption{Transmitted signal for $T_e = 0.25$s.}
	\label{TX_plot}
\end{figure}

\begin{figure}[!b]
	\centering
	\scalebox{0.5}{\includegraphics{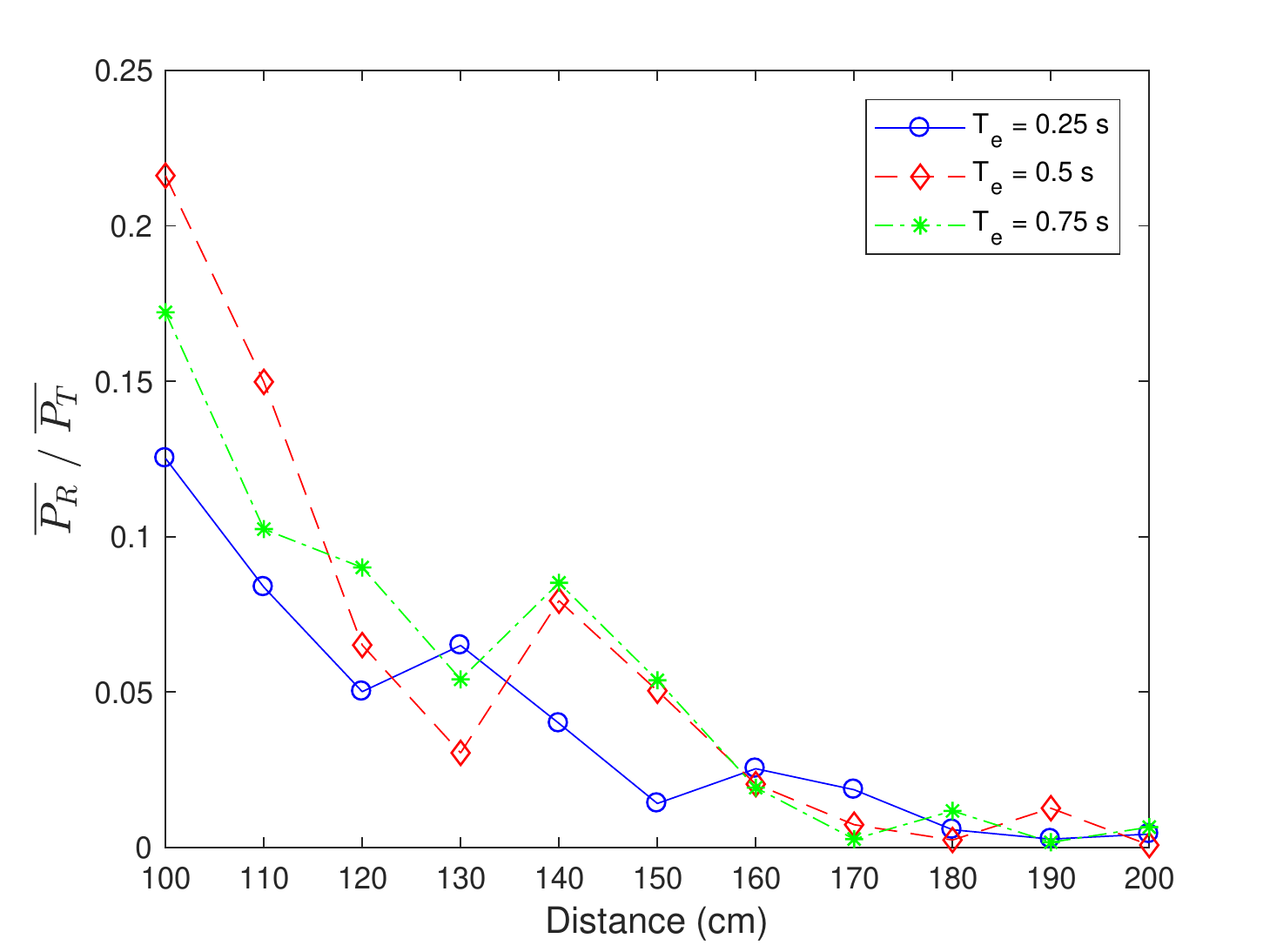}}
	\caption{$ \overline{P_R}/ \overline{P_T} $ vs. distance for different emission times.}
	\label{Power_plot}
\end{figure}

On the other hand, the transmitted power ($P_T$) is measured and averaged over $ 4 $ measurements for $ 3 $ different values of $T_e$ to obtain the average transmitted power ($\overline{P_T}$). The measurements are made by placing the sensor to a very close proximity of the TX, which  is assumed to generate a rectangular molecular pulse within $T_e$. The measured transmitted signal for 0.25 s emission time, is shown in Fig. \ref{TX_plot}. $\overline{P_T}$ values are $ 7.3598 $, $ 9.3666 $ and $ 11.0108 $ W for $ 0.25 $, $ 0.5 $ and $ 0.75 $ s emission times, respectively. Afterwards, $\overline{P_R}$ and $\overline{P_T}$ values are employed to derive a relation with the distance between the TX and RX. Fig. \ref{Power_plot} shows that there is a decreasing exponential relation among these parameters for three different emission times. 

The relation between $ \overline{P_R}/ \overline{P_T} $ and the distance is exploited to derive a function by using curve fitting techniques with the data illustrated in Fig. \ref{Power_plot}. The function defining the relation between  $ \overline{P_R}/ \overline{P_T} $ and the distance, is best fitted for the curve fitting model with the equation
\begin{equation}
\frac{ \overline{P_R}}{\overline{P_T} } = a_1 e^{b_1 d}, 
\label{Fit1}
\end{equation}
where $a_1$ and $b_1$ are the curve fitting parameters. The curve fitting is made by using nonlinear least squares method which minimizes the sum of squared errors. LM algorithm, which is explained in detail in Section \ref{ML}, is used to estimate the $a_1$ and $b_1$ coefficients in an iterative way. Furthermore, other algorithms such as the Newton or Gauss-Newton (GN) can be employed for curve fitting \cite{hansen2013least}. While the Newton algorithm has a more stable convergence than the GN algorithm, it has a cost of calculating second order derivatives. The GN algorithm is less complex, since it calculates first order derivatives instead of second order derivatives and thus, it has a faster (but less stable) convergence. The LM algorithm performs better than the GN algorithm in terms of stability for convergence with the cost of higher complexity.

\begin{figure}[H]
	\centering
	\includegraphics[width=0.3\textwidth]{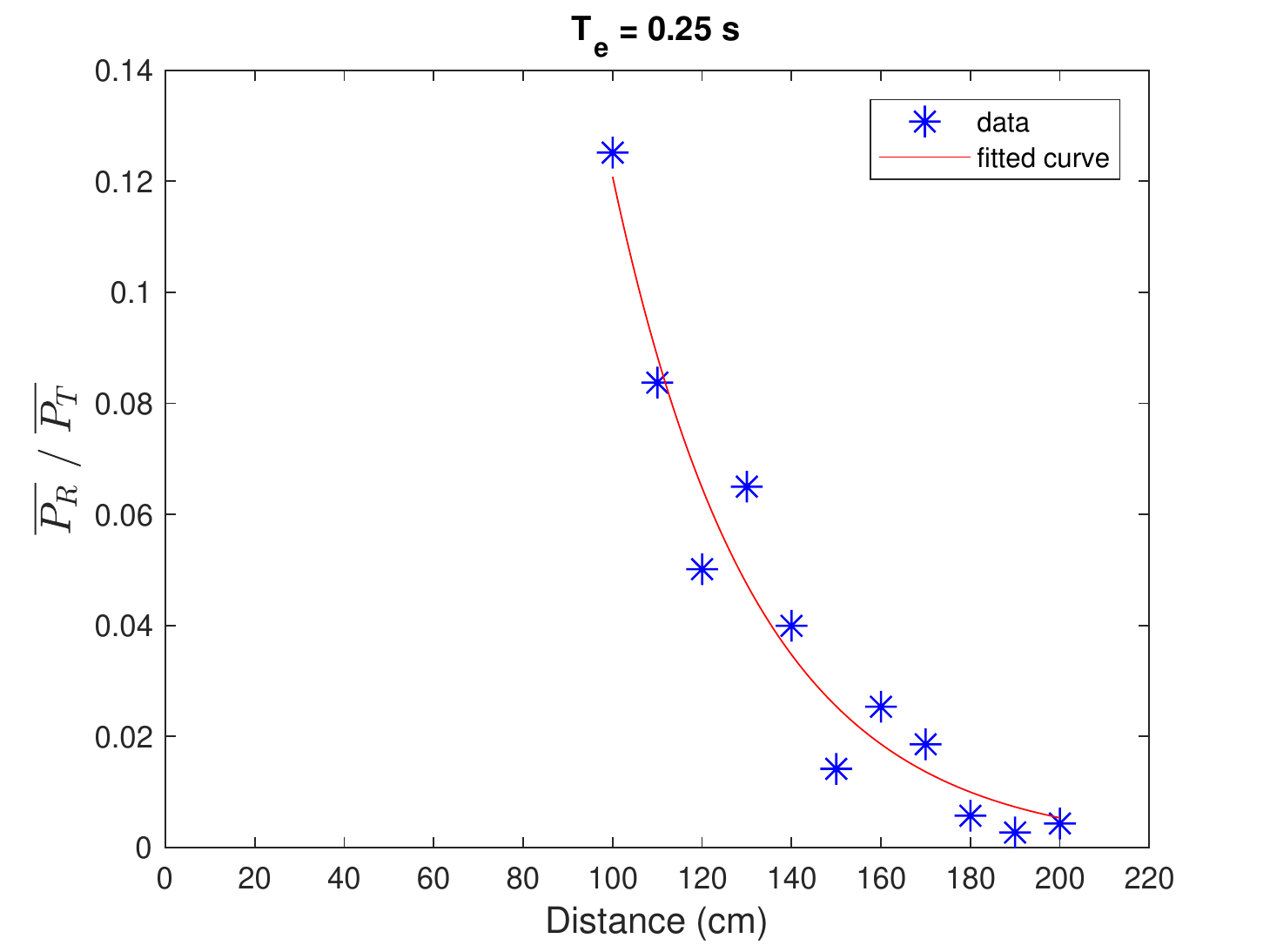}    
	\includegraphics[width=0.3\textwidth]{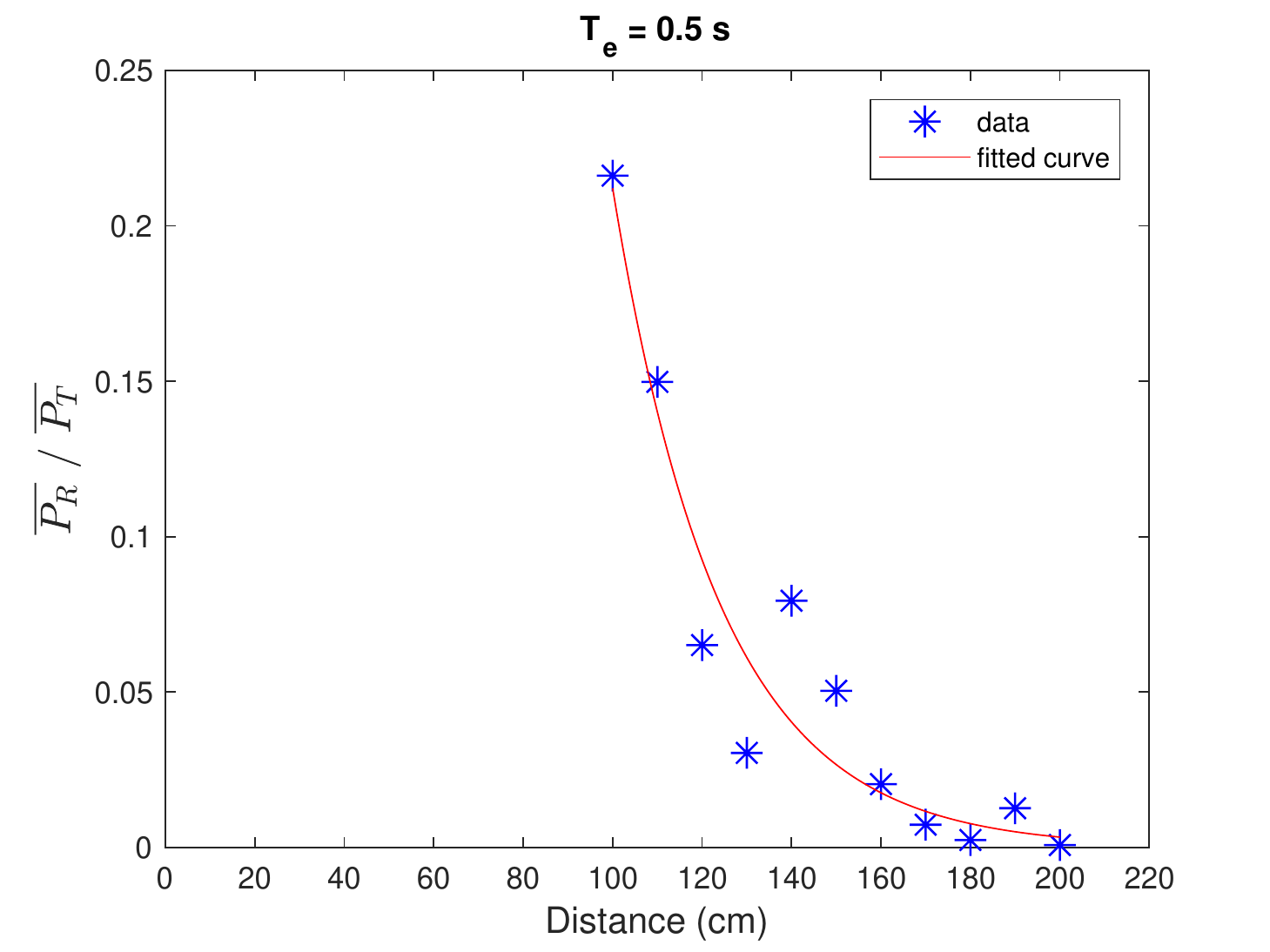}  
	\includegraphics[width=0.3\textwidth]{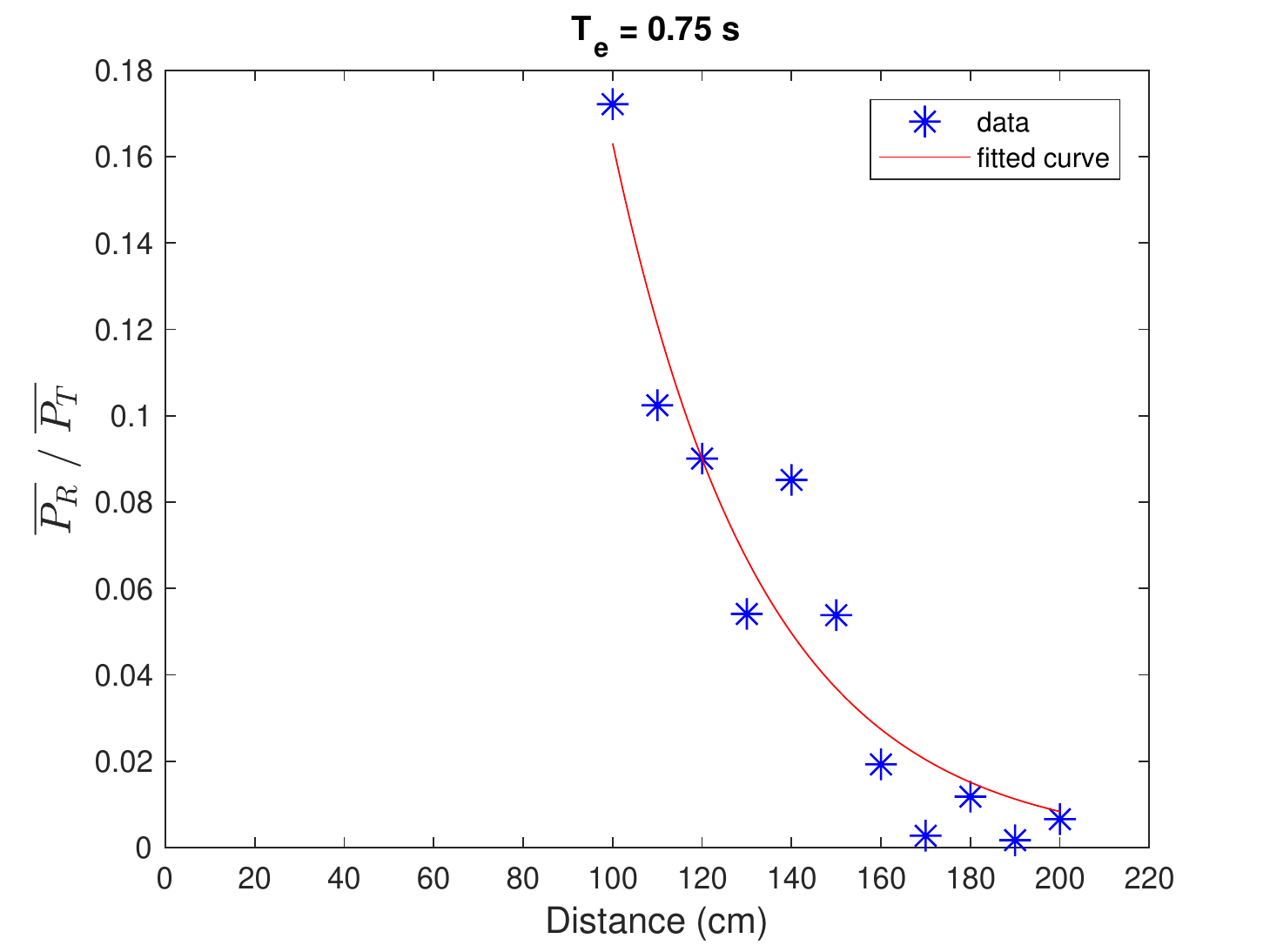} \\
	\scriptsize \hspace{0.6 in} (a) \hspace{1.8 in}  (b) \hspace{1.8 in} (c) \hspace{0.5 in}
	\caption{Measured $ \overline{P_R}/ \overline{P_T} $ vs. the distance and their fitted curves for (a) $T_e = 0.25$ s, (b) $T_e = 0.5$ s and (c) $T_e = 0.75$ s.}
	\label{P_fit}
\end{figure}

By using the LM algorithm, the measured data and their corresponding fitted curves are given in Fig. \ref{P_fit} and the calculated $a_1$ and $b_1$ parameters with the Root Mean Square Error (RMSE) values for  each $T_e$ are given in Table \ref{fp1}. As observed from the figure, the measured $\overline{P_R}/ \overline{P_T}$ decreases exponentially by the distance. This relation can be employed to estimate the distance by pulling out $d$ in (\ref{Fit1}) which is given by
\begin{equation}
\hat{d} = \frac{1}{b_1} ln\left(\frac{P_R}{P_T a_1}\right),
\end{equation}
where average power values are replaced with the measured power values and $ln(.)$ shows the natural logarithm.

\begin{table}[H]
	\caption{Curve fitting parameters for power based distance estimation.}
	\vspace{0.2cm}
	\centering
	\begin{tabular}{cccc}
		\hline
		$\mathbf{T_e}$	& $\mathbf{a_1}$ & $\mathbf{b_1}$ & $ \mathbf{RMSE} $ \\
		\hline
		$0.25$ s & 2.724 & -0.03116 & 0.0096 \\
		$0.5$ s & 13.4001 & -0.04146 &  0.0211\\
		$0.75$ s & 3.1888 & -0.02973 & 0.0171\\
		\hline
	\end{tabular}
	\label{fp1}
\end{table}

\subsection{Peak Time Based Distance Estimation}
\label{PTBDE}
\begin{figure}[H]
	\centering
	\scalebox{0.5}{\includegraphics{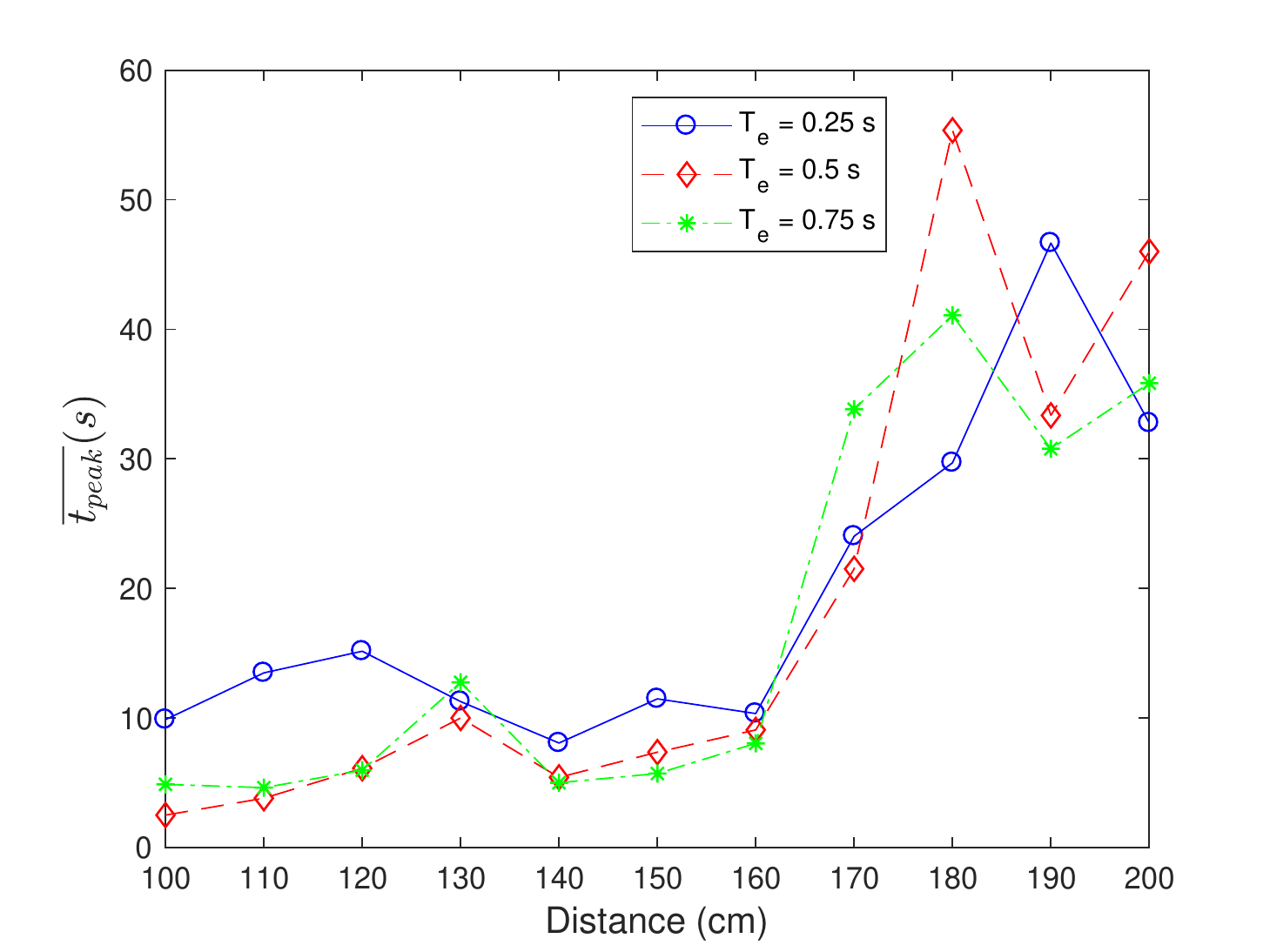}}
	\caption{Average peak time vs. distance for different emission times.}
	\label{t_plot}
\end{figure}

\begin{figure}[H]
	\centering
	\includegraphics[width=0.3\textwidth]{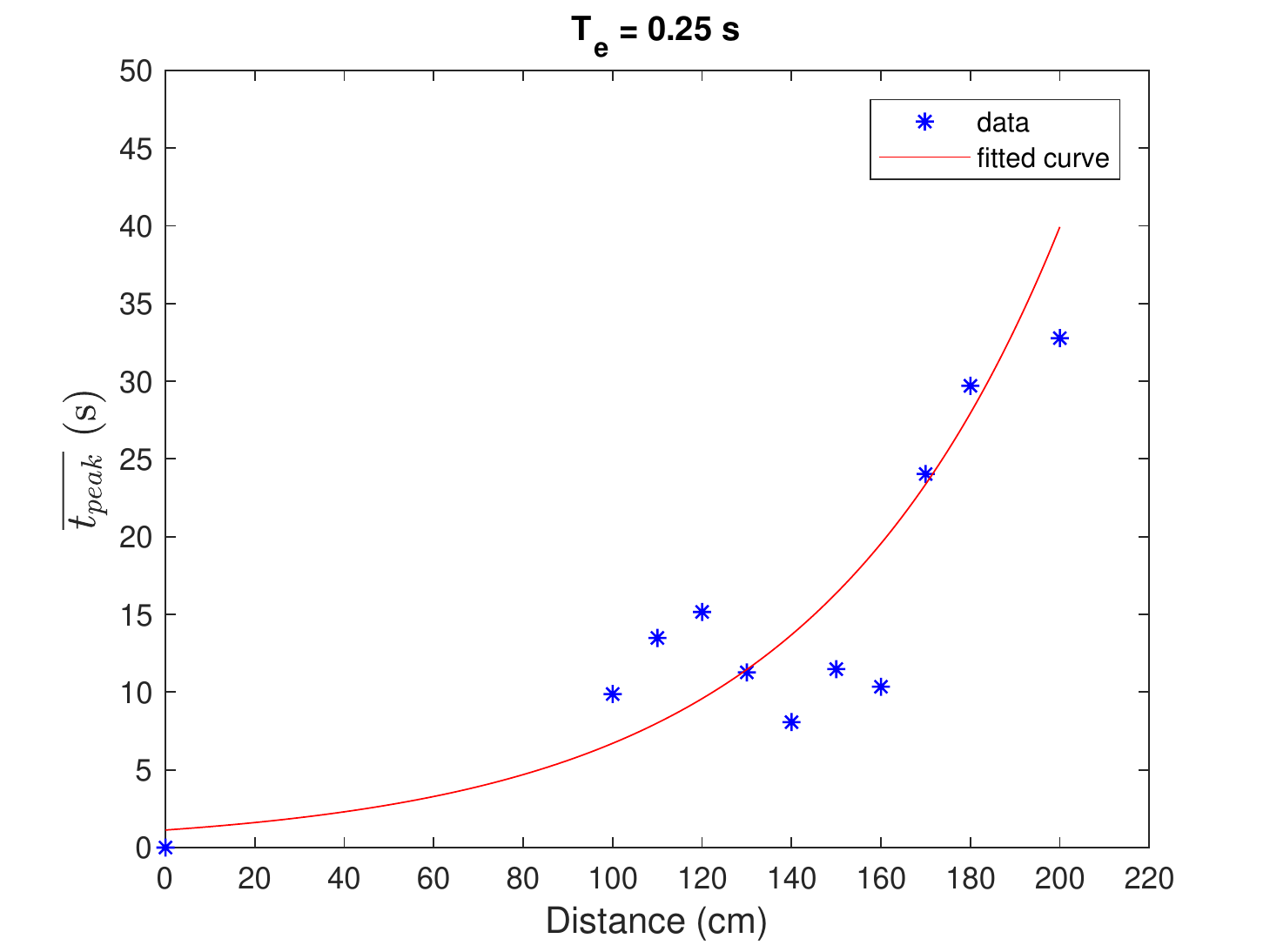}    
	\includegraphics[width=0.3\textwidth]{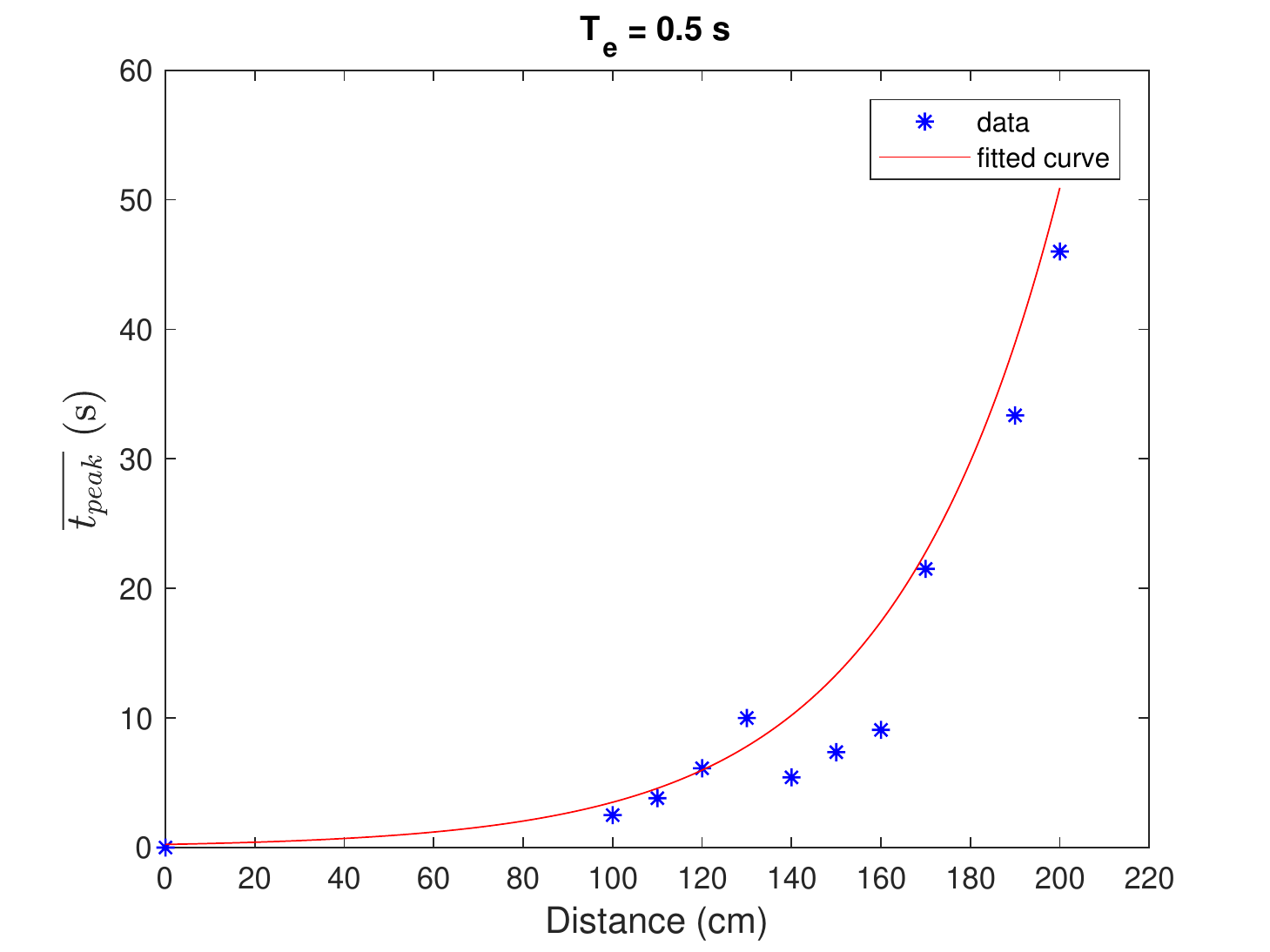}  
	\includegraphics[width=0.3\textwidth]{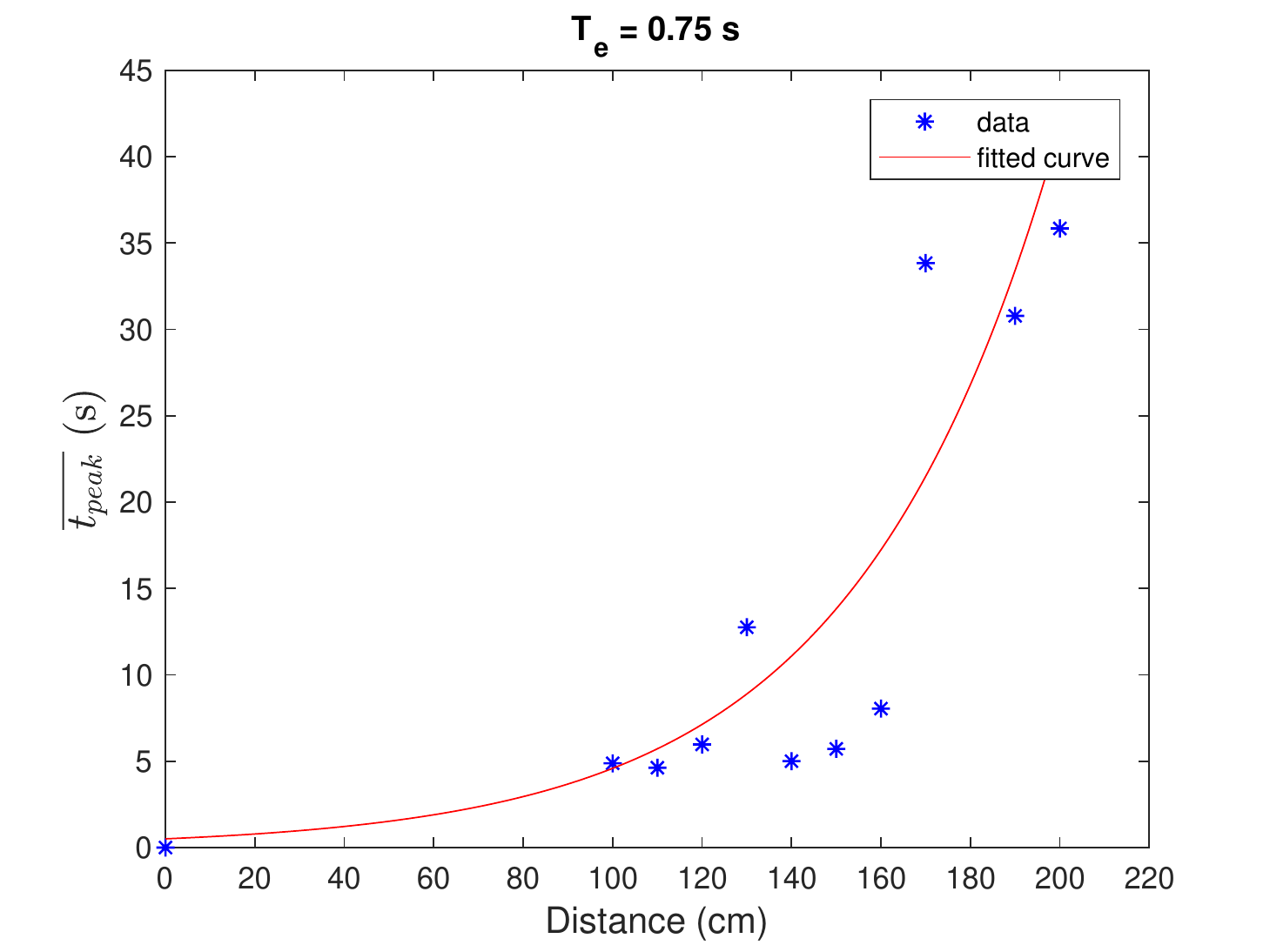} \\
	\scriptsize \hspace{0.6 in} (a) \hspace{1.8 in}  (b) \hspace{1.8 in} (c) \hspace{0.5 in}
	\caption{Measured average peak time vs. the distance and their fitted curves for (a) $T_e = 0.25$ s, (b) $T_e = 0.5$ s and (c) $T_e = 0.75$ s.}
	\label{t_fit}
\end{figure}

In this method,  similar to the power based distance estimation, the extracted $t_{peak}$ feature is averaged over the measured values for each distance. The averaging is performed separately for three different emission times as $ 0.25 $, $ 0.5 $ and $ 0.75 $ s. As observed in Fig. \ref{t_plot}, there is an increasing exponential relation between the average peak time and the distance. A function can be obtained by curve fitting with the formula given by
\begin{equation}
\overline{t_{peak}} = a_2 e^{b_2 d},
\label{fit2}
\end{equation}
where $ \overline{t_{peak}} $ is the average peak time, $a_2$ and $b_2$ are curve fitting parameters. An insertion is made to the data for $ \overline{t_{peak}} = 0$ at $d=0$ to obtain a more accurate curve. The same procedure is used for curve fitting as applied for power based distance estimation. According to (\ref{fit2}), the fitted curves are given in Fig. \ref{t_fit} for different emission times. By using these curve fitting models, their parameters and RMSE values are calculated as given in Table \ref{fp2}. Via the substitution of the measured peak time ($t_{peak}$) with $ \overline{t_{peak}}$, the relation between the average peak time and the distance are employed to derive a distance estimation model as given by 
\begin{equation}
\hat{d} = \frac{1}{b_2} ln\left(\frac{t_{peak}}{a_2}\right).
\end{equation}
\begin{table}[H]
	\caption{Curve fitting parameters for peak time based distance estimation.}
	\vspace{0.2cm}
	\centering
	\begin{tabular}{llll}
		\hline
		$\mathbf{T_e}$	& $\mathbf{a_2}$ & $\mathbf{b_2}$ & $ \mathbf{RMSE} $ \\
		\hline
		$0.25$ s & 1.1259 & 0.0178 & 6.6574 \\
		$0.5$ s & 0.2401 & 0.0268  & 9.1950\\
		$0.75$ s & 0.5045 & 0.0221 & 7.7602\\
		\hline
	\end{tabular}
	\label{fp2}
\end{table}

\subsection{Combined Distance Estimation}
The power based and peak time based distance estimation methods given in Section \ref{PBDE} and \ref{PTBDE}, respectively, can be combined to obtain a novel estimation method. When the ratio $P_R/P_T$ to $t_{peak}$ is written by using the curve fitting models in (\ref{Fit1}) and (\ref{fit2}), the following equation is obtained.
\begin{equation}
\dfrac{\frac{P_R}{P_T}}{t_{peak}} = \dfrac{a_1 e^{b_1 d}}{a_2 e^{b_2 d}} .
\label{fit3}
\end{equation}
By the manipulation of (\ref{fit3}), $d$ can be derived as
\begin{equation}
\hat{d} = \frac{1}{b_1-b_2} ln\left(\frac{P_R a_2}{P_T t_{peak} a_1}\right),
\end{equation}
which combines the power based and peak time based distance estimation methods. Next, the numerical results are given for all of the methods given up to now.

\section{Results and Analysis}
\label{Results_Comparison}
In this section, numerical results are presented for distance estimation to analyze the performance of ML methods given in Section \ref{ML} and the proposed data analysis based methods given in Section \ref{DEM}. First, the numerical results of the ML and data analysis based methods are discussed. Then, these methods are compared and an analysis about the experimental results are given.

\subsection{Numerical Results}
The aforementioned nine features, which are $ t_{low}$, $C_{low}$, $R$, $\Delta C$, $G$, $t_{peak}$, $C_{peak}$, $T_e$ and $E_R$, are given as the input for training the ML methods. Here, all of the features except $T_e$, are extracted as explained in Section \ref{Feature_Extraction}. Since $T_e$ is assumed to be known by the RX, it is also given as a feature. The Monte Carlo simulations are performed $10^5$ times for the ML algorithms to average the estimated distance values and their errors. The data are divided randomly as $ 70\% $ for training and $ 30\% $ for testing in MLR. In NNR, the data are randomly divided as $ 70\% $ for training, $ 15\% $ for validation and $ 15\% $ for testing. Furthermore, the hidden layer of the neural network has one node in the simulation model.

\begin{figure}[!h]
	\centering
	\scalebox{0.5}{\includegraphics{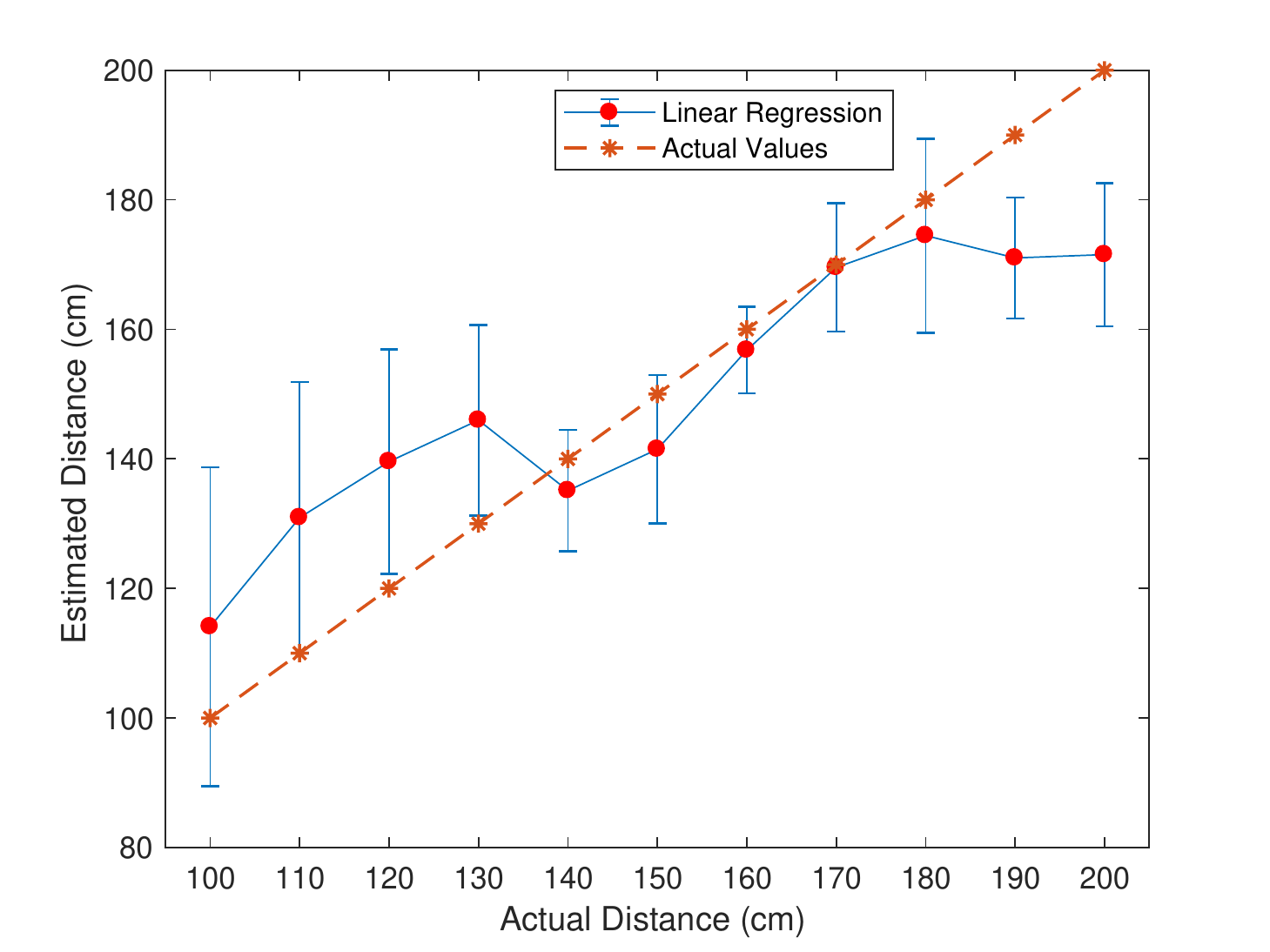}}
	\caption{Mean estimated distance and their standard deviations with linear regression for each distance.}
	\label{LR_plot}
\end{figure}

\begin{figure}[!h]
	\centering
	\scalebox{0.5}{\includegraphics{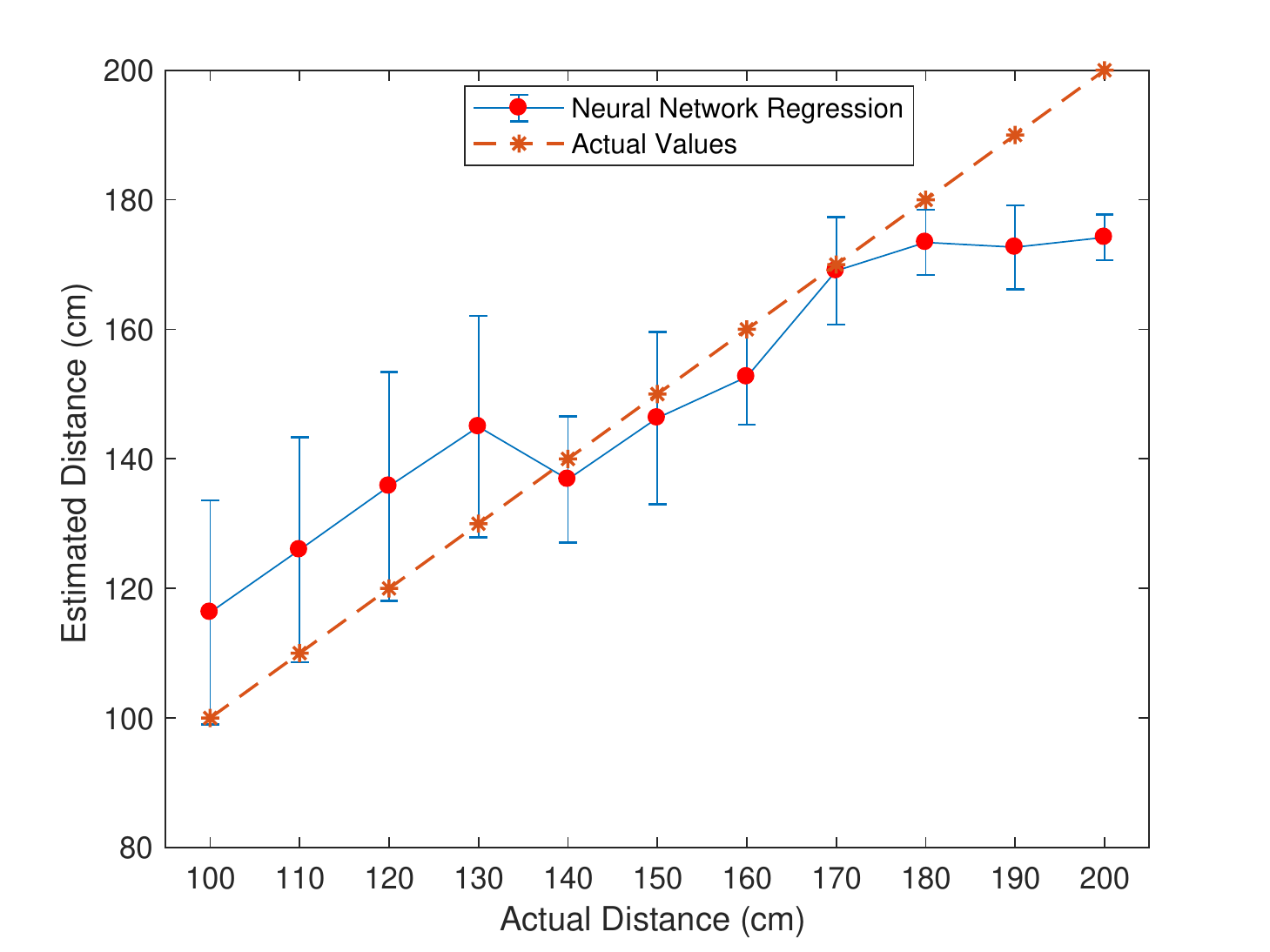}}
	\caption{Mean estimated distance and their standard deviations with neural network regression for each distance.}
	\label{NN_plot}
\end{figure}

The performance comparison of the ML methods  can be made for each distance. The mean estimated distances and their standard deviations with respect to actual distance are given in Fig. \ref{LR_plot},  and \ref{NN_plot} for MLR and NNR, respectively. The points at each distance show the mean estimated values and the vertical bars show their standard deviations. Moreover, a perfect estimation line is given to see how far the estimated values approach to the actual values. The reason for giving these figures is to see especially the standard deviations, i.e., to visualize how consistent the estimated values are. The NNR has slightly better results than MLR for the mean estimated values and standard deviations. However, since the difference between MLR and NNR results is small, it can be deduced that  the relationship  between the features used as input and the distance is highly linear.  The cost of the NNR method is its complexity and longer time requirement. Hence, estimating the distance with MLR can be considered as an efficient ML method due to its simplicity and performance for our practical scenario and extracted features. 
\begin{figure}[!ht]
	\centering
	\scalebox{0.55}{\includegraphics{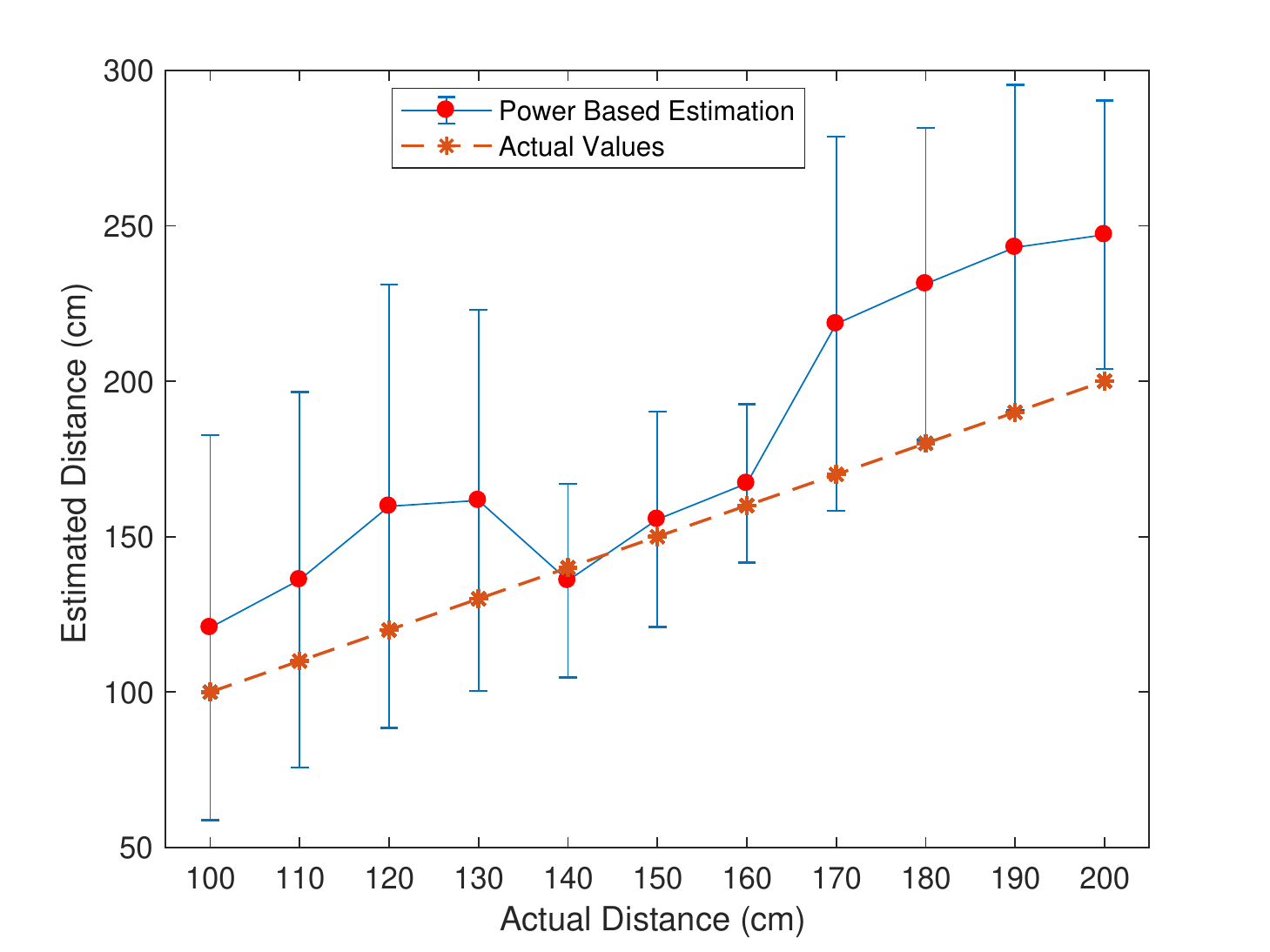}}
	\caption{Mean estimated distances and their standard deviations for power based distance estimation.}
	\label{pbe_plot}
\end{figure}
\begin{figure}[!ht]
	\centering
	\scalebox{0.5}{\includegraphics{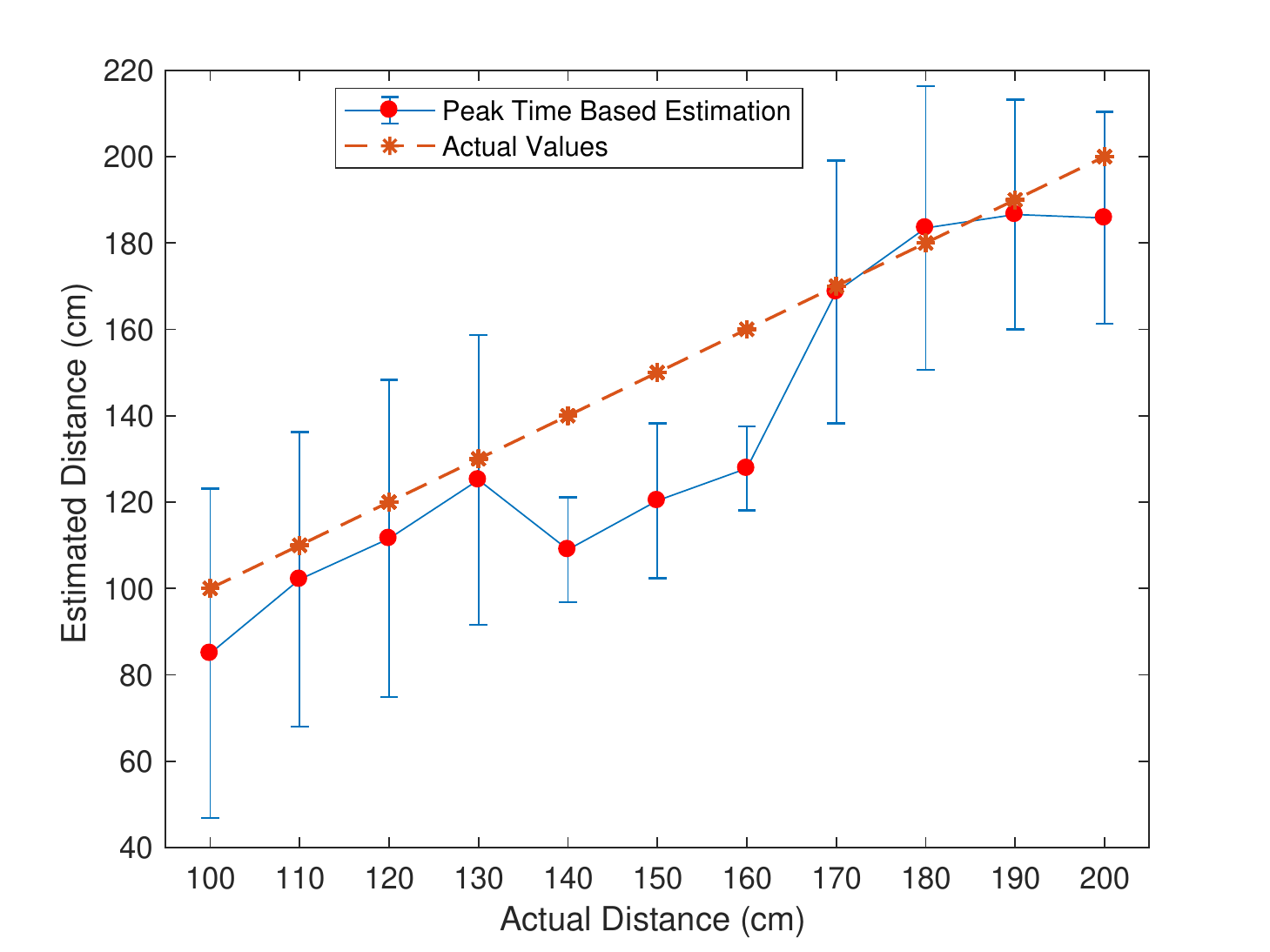}}
	\caption{Mean estimated distances and their standard deviations  for peak time based distance estimation.}
	\label{ptbe_plot}
\end{figure}
\begin{figure}[!hbt]
	\centering
	\scalebox{0.5}{\includegraphics{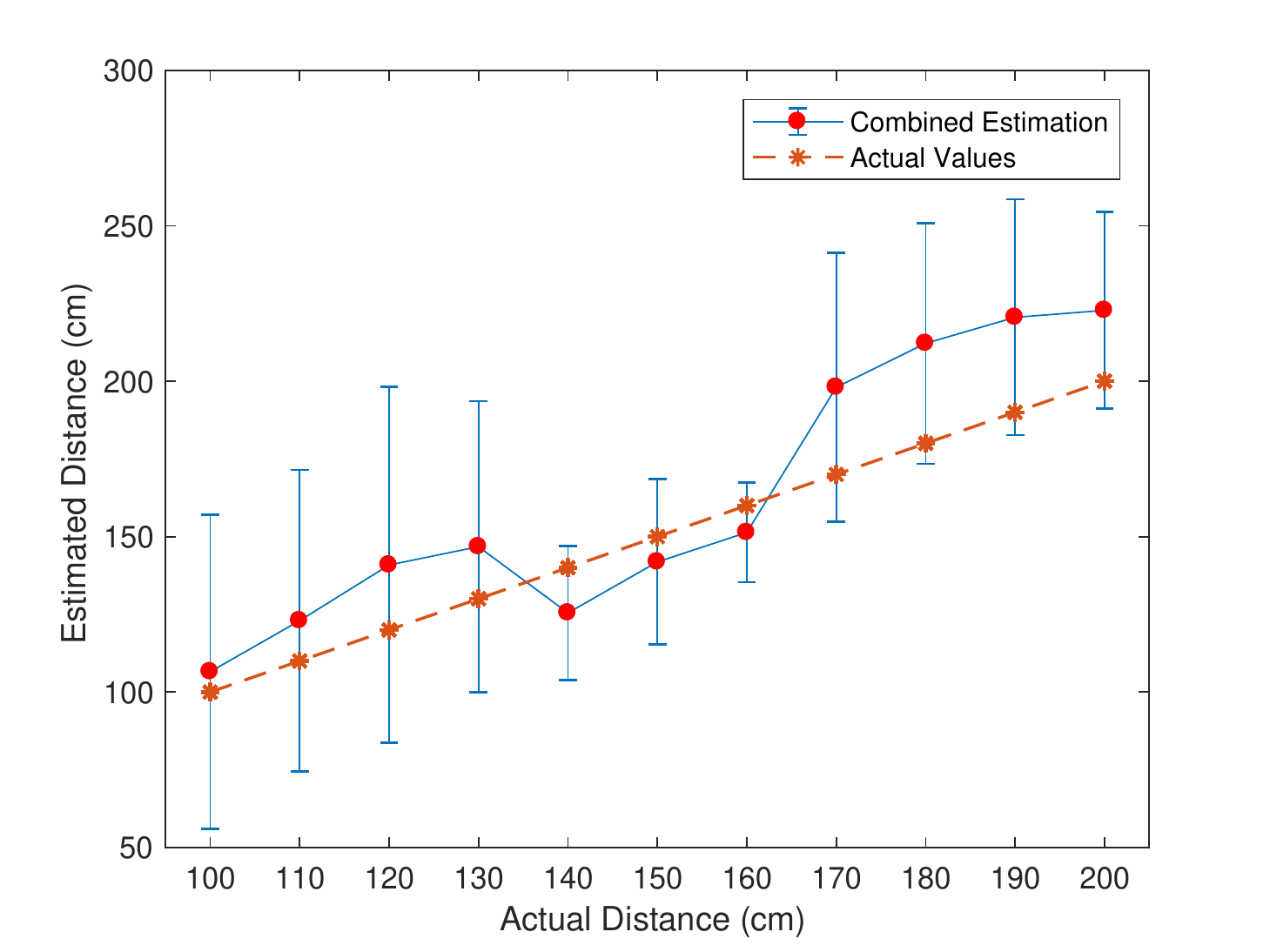}}	
	\caption{Mean estimated distances and their standard deviations for combined distance estimation.}
	\label{ce_plot}
\end{figure}

The performance evaluation of the proposed data analysis based distance estimation methods given in Section \ref{DEM}, are made by using the features of the measured signals. The mean estimated distance values and their standard deviations are shown in Figures \ref{pbe_plot}, \ref{ptbe_plot} and \ref{ce_plot} for power based, peak time based and combined estimation methods, respectively. The values for every actual distance on these figures show the mean (dots) and standard deviation (vertical bars) values of the estimated distance over 30 measurements. Due to the smaller number of features employed in data analysis based methods, the standard deviations are relatively higher, when it is compared with the ML methods. There can be some unused parameters in the data analysis based methods to estimate the distance more accurately with respect to the ML methods. Next, all of the results is given in a comparable format to evaluate the performance of the distance estimation methods given in this paper.

\subsection{Comparison}
In Table \ref{Results}, the Root Mean Square Error ($\rho$) values are given for each method, which is defined by
\begin{equation}
\rho = \sqrt{\frac{1}{M}\sum_{i=1}^{M} (\hat{d_i}-d_i)^2},
\end{equation}
where $M$ is the number of the samples of the test data. In data analysis based methods, $M$ is taken as $ 330 $ by using whole data samples. In ML methods, $M$ represents the number of the samples of the test data as defined in the beginning of this section and $\rho$ is averaged over $10^5$ Monte Carlo simulation trials. The results for ML methods are much better than the data analysis based methods. However, the cost of the superiority of the ML methods emerge as time complexity which means that the ML methods require more time and processing power for distance estimation calculations due to the longer equations used in these methods. NNR has the highest complexity among the proposed methods. Furthermore, the training phase of the ML methods can restrict real-time applications due to this complexity. Therefore, there is a trade-off between the complexity and performance for the ML and data analysis based methods, which needs to be optimized for the implementation of an application. On the other hand, the fact that the data analysis based methods have less time complexity, makes the implementation of these methods easier than the ML methods. 
\begin{table}[!hbt]
	\caption{Distance estimation performance for the ML and data analysis based methods.}
	\vspace{0.2cm}
	\centering
	\begin{tabular}{cc}
		\hline
		\textbf{Method}	& \textbf{RMSE} \\
		\hline
		Linear Regression & 21.4470  \\
		Neural Network Regression & 20.1157 \\ 
		Power Based Estimation & 62.3077  \\
		Peak Time Based Estimation & 33.2434 \\
		Combined Estimation & 44.3157 \\
		\hline
	\end{tabular}
	\label{Results}
\end{table}

\begin{figure}[!htb]
	\centering
	\scalebox{0.6}{\includegraphics{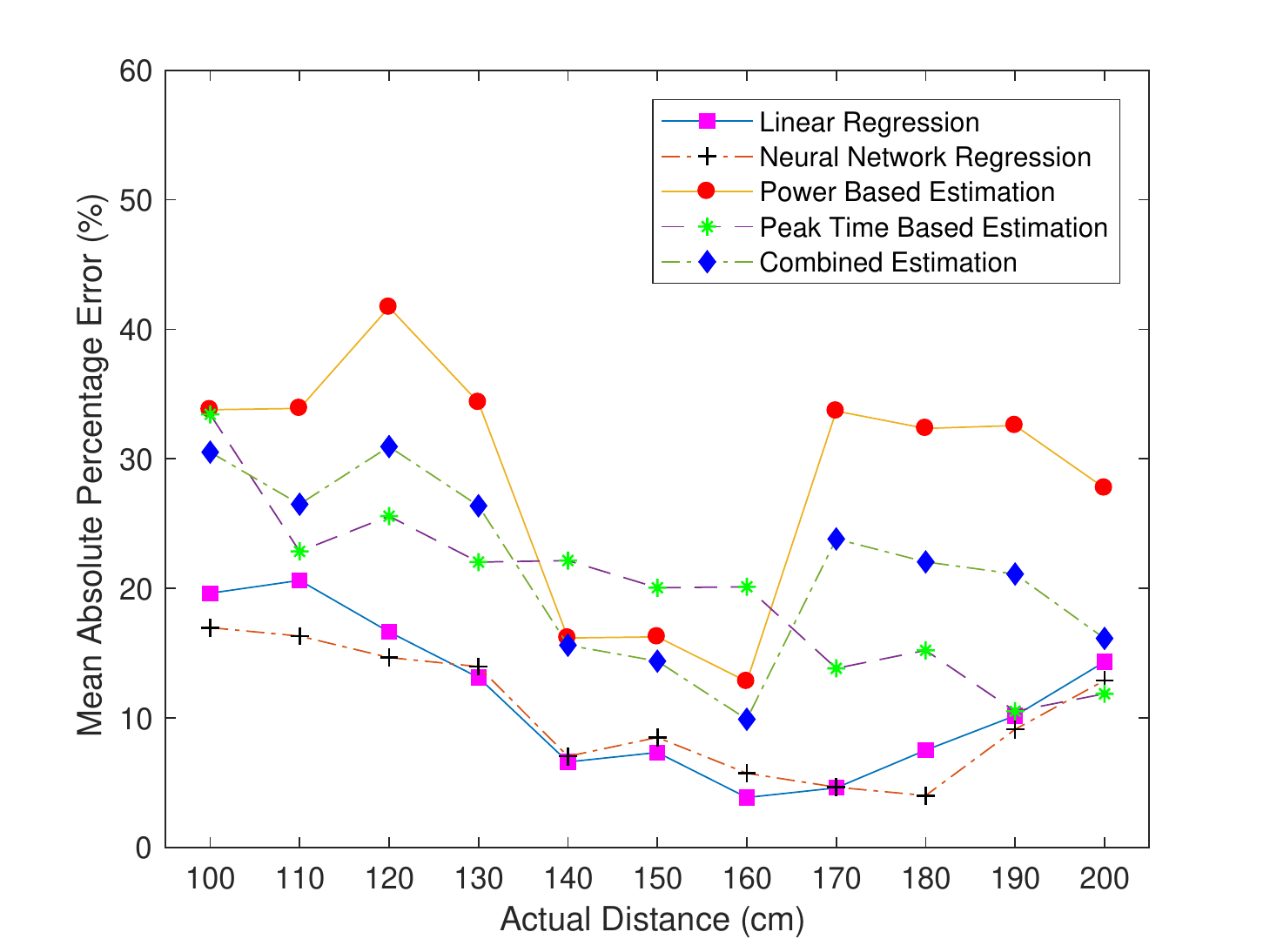}}
	\caption{Mean absolute percentage errors vs. actual distance for the ML and data analysis based methods.}
	\label{All_perc_plot}
\end{figure}

While the distance values increase, the same absolute error deviation has different meanings according to the actual distance and the error values need to be normalized to determine which method is better. Hence, mean absolute percentage error (MAPE) is preferred as the performance metric rather than the absolute error deviation. For each actual distance, the MAPE ($\epsilon$) values shown in Fig. \ref{All_perc_plot} are calculated according to the formula as given by 
\begin{equation}
	\epsilon = \frac{100}{M} \sum_{i=1}^{M} \frac{|\hat{d_i}-d|}{d}.
	\label{MAPE}
\end{equation}

In our practical scenario, our observations show that some of the droplets fall to the ground, before they arrive to the RX. The droplets are sprayed from the TX as sufficiently large droplets to be affected by the gravity \cite{de2017investigation}. Moreover, there are other effects such as boundary conditions in the diffusion process and the initial velocity of the molecules. Since ML has more parameters to model these effects analytically, they perform better than the data analysis based methods except the peak time based estimation method at long distances. In general, the peak time based estimation provides the best performance among the data analysis based methods. However, the peak time based estimation method has the worst performance between $140$ and $160$ cm due to the non-linearity of the sensor as detailed in Section \ref{Analysis}.

Unlike the other methods, peak time based estimation has a nearly linear estimation performance and its MAPE decreases, as the distance increases. At $ 190 $ and $ 200 $ cm, it has very close values with ML methods and even performs better than the ML methods at $ 200 $ cm. This result shows us that peak time based estimation can be applied to longer distances. Furthermore, it reveals that the peak time provides more accurate distance estimation, as the RX gets farther from the TX. Hence, the distance can be estimated with more sensitive sensors for farther distances.

In order to determine the optimal distance estimation method for a practical scenario, the processing capacity of the RX is essential. If the RX does not have the capacity to implement ML algorithms, then the peak time based estimation method can be used with a cost of slightly higher error than ML methods. Otherwise, MLR can be used as a more efficient method between ML methods due to its lower complexity and high accuracy. NNR has the best performance among all the proposed methods in terms of accuracy, although it imposes the highest complexity. Next, our observations and numerical results are analyzed to explain the phenomena behind the propagation of the molecules and its reception by the RX.

\subsection{Analysis of the Results} \label{Analysis}
\begin{figure}[!htb]
	\centering
	\scalebox{0.5}{\includegraphics{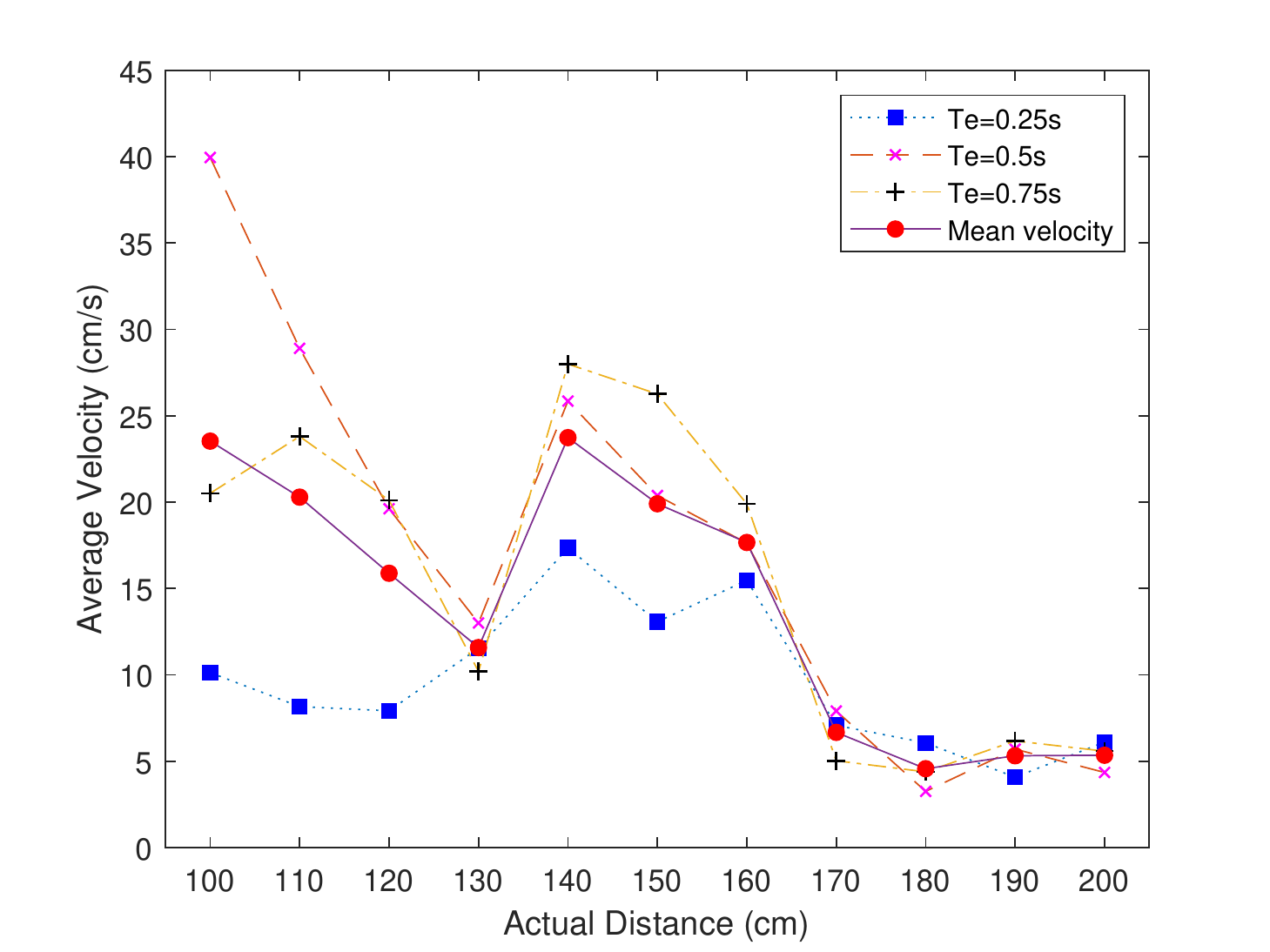}}
	\caption{Average velocity profile of the transmitted molecules with respect to the actual distance.}
	\label{Velocity}
\end{figure}

The lowest MAPE values are between $ 140 $ and $ 180 $ cm for ML methods and between $ 140 $ and $ 160 $ cm for data analysis based methods, except the peak time based estimation. The reason why the estimated values are better within the interval $ 140 $ and $ 160 $ cm, lies in the measured signals. This can be more clearly understood with the average velocity profile of the transmitted molecules in the medium as shown in Fig. \ref{Velocity}. In order to calculate the average velocity of the molecules, it is assumed that the majority of the molecules arrive at the RX, when the received molecule concentration is at its peak point. Although some of the molecules reach the RX before or after this peak point, it is reasonable to choose $t_{peak}$ as the average arrival time of the molecules. Hence, the average velocity of the molecules is defined as the ratio of the distance to the time duration at which the peak concentration is read from the sensor. In Fig. \ref{Velocity}, average velocities are obtained according to this definition for each $T_e$ and also the mean velocity is given as the average of the velocity values given for  three different values of $T_e$. 
\subsubsection{The Effect of the  Sensor's Non-linearity}
\begin{figure*}[!htb]
	\centering
	\includegraphics[width=0.42\textwidth]{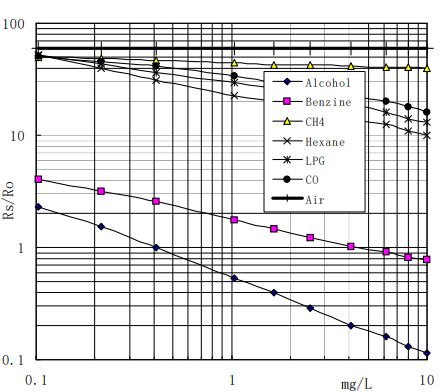}    
	\includegraphics[width=0.51\textwidth]{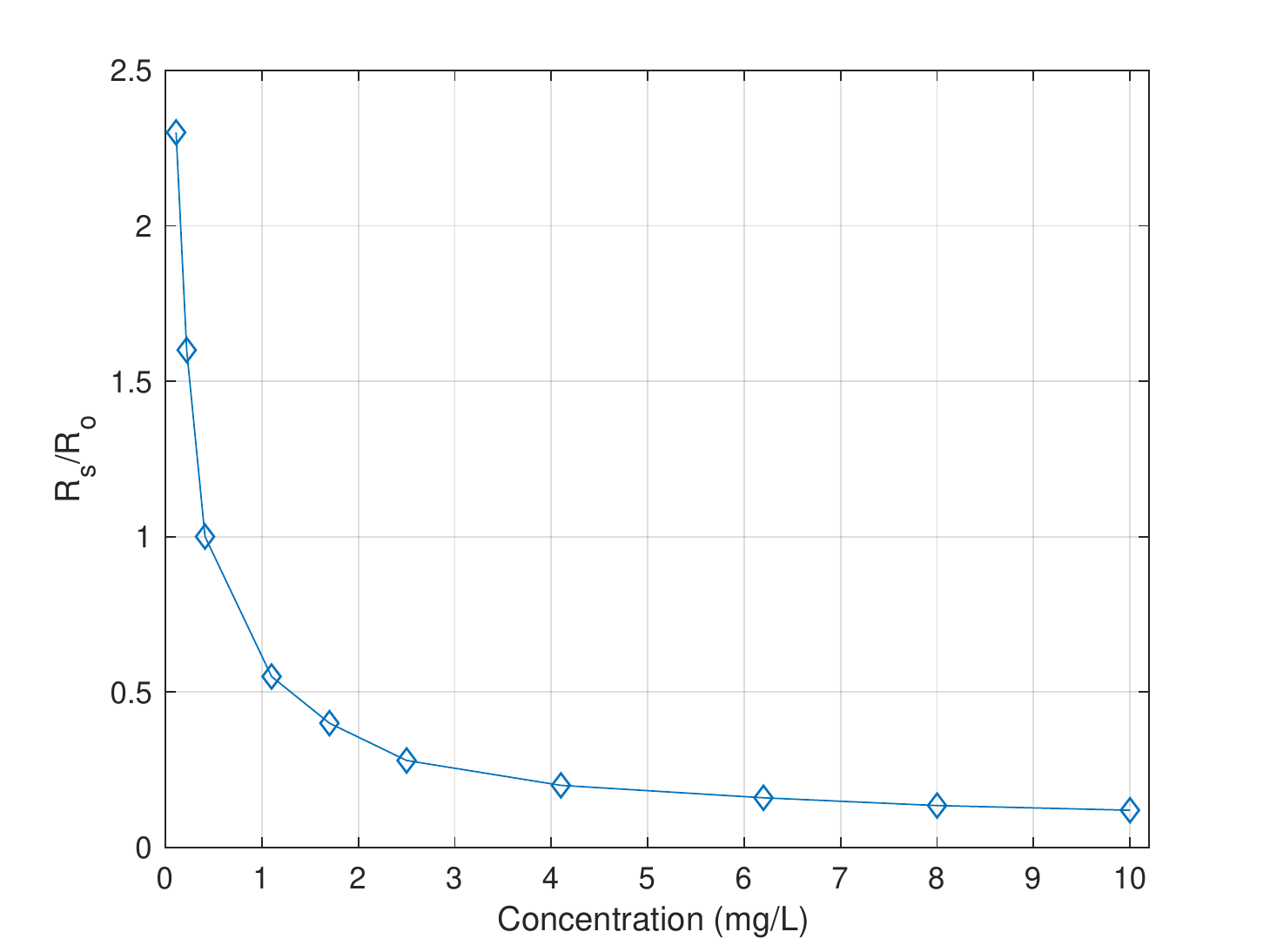} \\
	\scriptsize \hspace{-0.2 in}  (a) \hspace{2.85 in} (b)
	\caption{Sensitivity characteristics of the MQ-3 sensor ($R_s/R_o$ vs. Concentration)  \cite{MQ3} (a) for different gases in logarithmic scale  (b) for alcohol in linear scale (plotted using original values from the datasheet).}
	\label{MQ3}
\end{figure*}

The larger velocities of the molecules for the measurements between $ 140 $ and $ 160 $ cm can be related to the non-linearity of the sensor characteristics as given in Fig. \ref{MQ3}. In this figure, $R_s$ is the sensor resistance which changes according to the molecule concentration and $R_o$ is the sensor resistance measured at the concentration level $ 0.4 $ mg/L. The sensitivity of the sensor is shown for different gases in logarithmic scale in Fig. \ref{MQ3} (a) \cite{MQ3}. The values in this figure for alcohol is plotted in linear scale in Fig. \ref{MQ3} (b) to see the non-linearity of the sensor more clearly for our case. The MQ-3 sensor makes more sensitive measurements in lower concentrations with respect to higher concentrations. Due to this fact, the sensor can have errors for the measurements at lower distances.  Another possible explanation to understand the phenomena behind the motion of the molecules in our scenario, can be made by employing the dynamics of the fluid which is considered as follows.
\subsubsection{The Effect of Factors Related to Fluid Dynamics}
In ideal conditions where the molecules only make an isotropic random walk without being exposed to any force, the diffusion process  can be used to model the movement of the molecules. However, the molecules come out of the TX as liquid droplets with an initial velocity in our case. Thus, several processes concerning the dynamics of the fluid  can occur and the diffusion process is not sufficient to understand the motion of the molecules. Therefore, the analysis is needed to be made in terms of fluid dynamics. 

\begin{figure}[!htb]
	\centering
	\scalebox{0.55}{\includegraphics{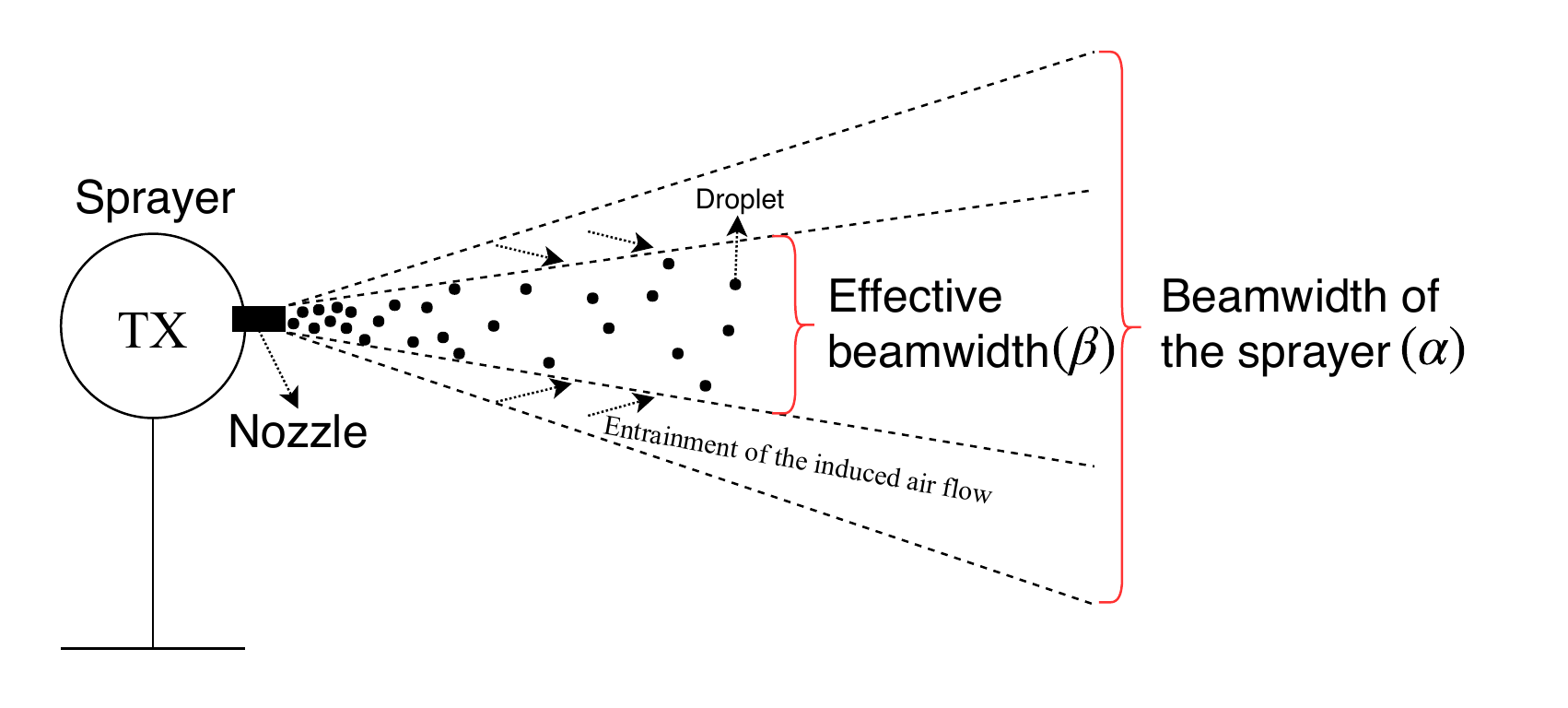}}
	\caption{Beamwidth of the sprayer (TX) and the effective beamwidth.}
	\label{beam}
\end{figure}

When a liquid is sprayed into the air, the droplets initially come out of the sprayer with a certain angle which is the beamwidth (spray angle) of the sprayer as shown in Fig. \ref{beam} with $\alpha$. The interaction between the transmitted droplets and the air molecules leads to a contraction in the effective beamwidth, which is defined as the angle of the propagating droplets as given in Fig. \ref{beam} with $ \beta $. This interaction is detailed as follows.

As the droplets move in the air due to the force applied by the sprayer, they gain a momentum, which can be defined as the product of the liquid's mass and velocity. According to the law of momentum conservation, when two objects are collided, their total momentum does not change before and after the collision \cite{kleppner2013introduction}. The collisions of the droplets with the air molecules cause aerodynamic drag which can be defined as the air force resisting the movement of the object \cite{anderson2016fundamentals}. The ethyl alcohol droplets slow down due to the aerodynamic drag. Besides, the momentum lost by the droplets is gained by the air molecules due to the momentum conservation. The momentum exchange among the droplets and air molecules generates an air flow towards the center of the beamwidth along the horizontal axis  \cite{rothe1977aerodynamic}. This flow drags the droplets into the inner regions from the outer regions. Namely, the angle at which the droplets propagate is reduced relative to the initial $\alpha$ angle, and this leads to a contraction. This contraction of the droplets' propagation angle facilitates turbulent flows which is detailed as follows \cite{ghosh1994induced}. 

The flows can be classified into two types as \textit{laminar} flow, which has parallel and regular streamlines, and \textit{turbulent} flow, which has random and irregular streamlines \cite{anderson2016fundamentals}. Even if the effect of the turbulent flow can be observed, its mathematical modeling is still a big problem due to its complicated nature. The flow type of the fluid depends on the Reynolds number ($Re$). In our case, $Re$ is given as \cite{ghosh1994induced},
\begin{equation}
Re = \frac{2 |V_l - V_a| a}{\nu_a},
\label{Re}
\end{equation}
where $a$ is the radius of the spherical droplet, $\nu_a$ is the kinematic viscosity of the air, $V_l$ and $V_a$ are the velocities of the liquid and airflow. $Re$ influences the type of viscous flows. As $Re$ increases, it is more likely to have a turbulent flow in the medium. There is a critical value of $Re$ to determine the flow type \cite{avila2011onset}. Above this critical value, which depends on the parameters of the transmitted molecules and the medium, the flow becomes turbulent. In order to determine where there is a turbulent flow according to (\ref{Re}), the radii of the liquid droplets, the velocities of the liquid droplets and air molecules are needed to be measured, which is beyond the scope of this paper. High-speed photography techniques or Phase Doppler Anemometry (PDA) can be used to measure the liquid droplet size and velocities. In \cite{begg2009vortex}, the movement of the liquid gasoline fuel droplets after they come out of a sprayer (injector) is analyzed. After the spraying operation, vortex ring-like structures, which include turbulent flows, are observed experimentally for the liquid droplets. As they move away from the sprayer, the average liquid droplet size gets smaller \cite{de2017investigation}. Brownian motion does not have an important effect, if the droplet is large enough. Hence, Brownian motion is more effective, after the velocity of the droplets decreases to a more steady value. The findings and results in \cite{begg2009vortex} can be used to analyze the movement of the droplets for our scenario, since the densities of the gasoline fuel and ethyl alcohol have close values \cite{mavrik2014comparison}. 

\begin{figure}[!h]
	\centering
	\scalebox{0.55}{\includegraphics{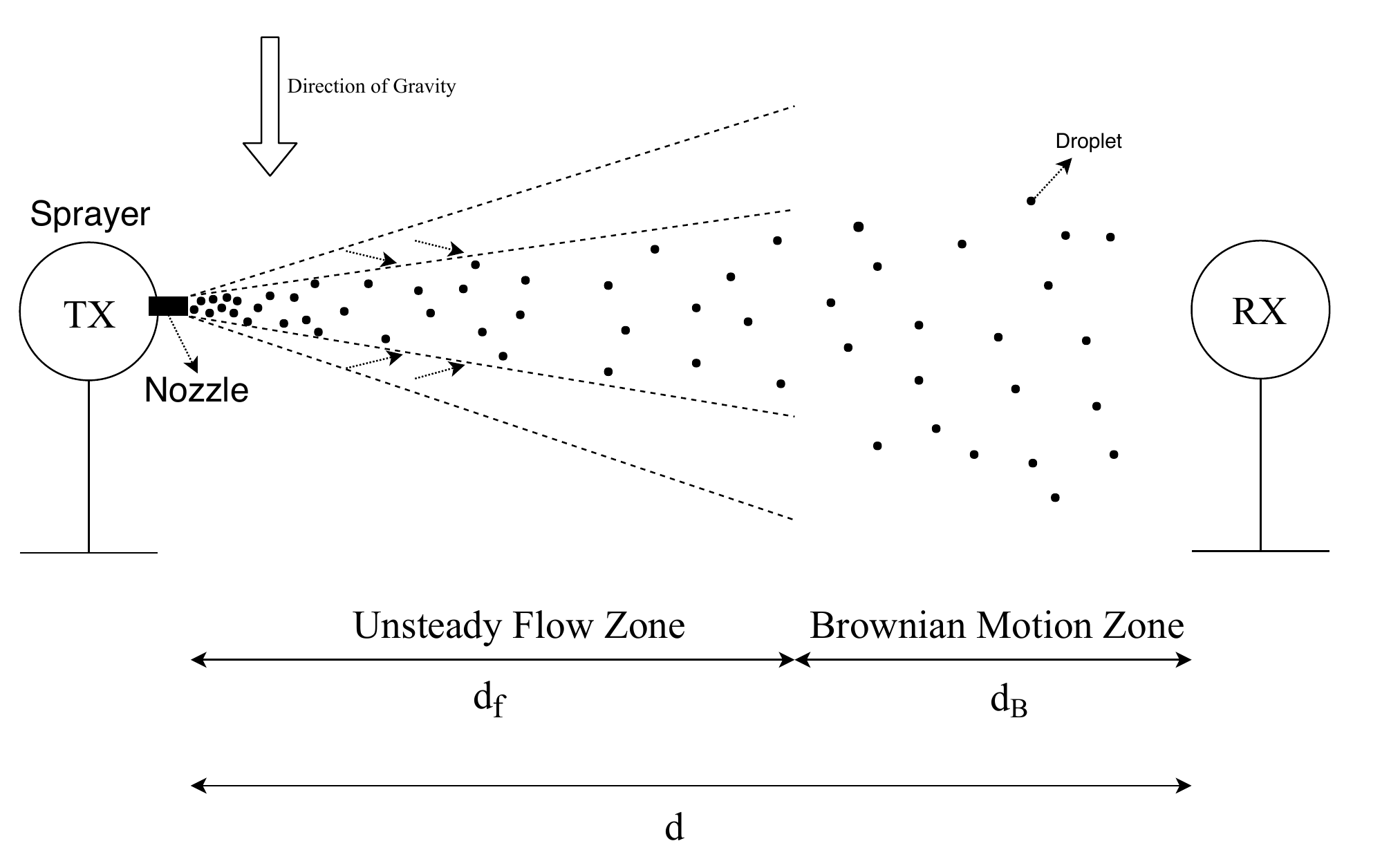}}
	\caption{The trajectory of the transmitted molecules between the TX and the RX.}
	\label{Flows}
\end{figure}

In the vicinity of the nozzle of the TX, the flow of the droplets can be defined as unsteady flow, since the velocity of the droplets are not constant with respect to time due to the interaction with the air molecules \cite{munson2009brief}. Moreover, if a constant flow is stopped suddenly, the flow shows unsteady characteristic \cite{munson2009brief}. Similarly in our case, since the TX suddenly stops its transmission at the end of $T_e$, it can be deduced that the flow in our case becomes an unsteady flow. As soon as molecules are emitted,  they are affected  by the Brownian motion which can be neglected in the region where  molecules are entrained by the higher velocities of unsteady flows such as turbulent flows. As the droplets move away from the TX, the aerodynamic drag decreases their velocities until a point where they start to move with only Brownian motion.  As the result of these movements and findings in \cite{begg2009vortex}, the transmitted droplets can be expected to have a trajectory as illustrated in Fig. \ref{Flows}. Here, the motion of the droplets is divided into two zones as the unsteady flow zone and Brownian motion zone. In the first zone, the droplets where they travel a distance $d_f$, are affected by the unsteady flows such as turbulent flows.  The droplets are exposed to random fluctuating aerodynamic drag forces due to the turbulent flows \cite{maxey1987gravitational}. When the effect of the flows ends, the droplets start to move with Brownian motion where they travel a distance $d_B$ until the RX. Although gravity affects the motion of the droplets in both zones, its effect is greater in the unsteady flow zone, since the size of the droplets is larger. By using the trajectory in Fig. \ref{Flows}, the velocity profile in Fig. \ref{Velocity} can be explained as follows. For the measurements between $ 100 $ and $ 130 $ cm, the non-linearity of the sensor can cause more measurement errors due to the lower sensitivity in higher concentrations. In addition to this,  the average distance traveled in the unsteady flow zone  can be larger for the measurements between $ 140 $ and $ 160 $ cm  than the average distance traveled in the unsteady flow zone for the measurements at other distances due to  effects of the unsteady flows. Thus, the droplets move with larger  velocities which cause shorter arrival time to the RX and a longer $ d_f $ than the measurements at other distances. Consequently, these factors such as non-linear characteristic of the sensor and  unsteady flows jointly affect the performance for distance estimation.

As given by the analysis about the unexpected average velocities of the molecules, their motion for macroscale practical scenarios can be modeled by exploiting fluid dynamics. The channel modeling and parameter estimation with the fluid dynamics point of view are open research issues for practical MC scenarios. Next, the role of emission time, i.e., $ T_e $, is examined for distance estimation by using the trajectory given in Fig. \ref{Flows}.

\subsubsection{The Effect of the Emission Time}
Fig. \ref{Velocity} for short distances up to 130 cm, the average velocity of the molecules for $T_e = 0.5$ s are greater than the average velocity of the molecules for $T_e = 0.25$ s. However, for $T_e = 0.75$ s, the average velocity is less than the molecules transmitted with $T_e = 0.5$ s (except the nearly equal value at 120 cm). This result shows that there is an optimum $T_e$ for the fastest propagation of the molecules according to the distance. Furthermore, it shows that sending more molecules than this optimum $T_e$ can decrease the average velocity of the molecules. When $T_e$ is larger  than the optimum $T_e$, a higher molecule concentration occurs in the unsteady flow zone. This increases the collision rate of the molecules and thus, reduces their velocity. For distances between 140 and 160 cm, the average velocity is directly proportional with $T_e$ due to the above-mentioned longer average distance traveled in the unsteady flow zone. For long distances greater than 170 cm, $T_e$ does not significantly affect the average velocity of the molecules. Hence, it can be deduced from the results that an appropriate $T_e$ should be chosen for a faster propagation according to the distance. More experimental research is needed for the derivation of the optimum $T_e$, which is left as a future work.

\section{Conclusion}
\label{Conclusion}
In this paper, five distance estimation methods are proposed for a practical macroscale MC system. The existing two ML methods, which are linear  and neural network regression, are applied in distance estimation for the first time. In order to use these methods, an experimental setup was established and received signals were recorded. A novel feature extraction algorithm is proposed to generate training and test data from the measured signals for the ML methods. By analyzing these generated data, three novel methods for distance estimation are proposed and compared with the applied ML methods. The ML methods result better than data analysis based methods, with NNR being the best method. However, MLR has a very close performance to NNR, which shows that the input features such as peak time, gradient of the received signal during the transition from the noise floor up to the peak time and the received energy of the signal, have a linear relation with the distance. MLR can be an efficient way for distance estimation with high accuracy and low complexity as a ML method. Furthermore, data analysis based methods perform worse, but are not as complex as ML methods. Especially, the peak time based estimation performs close to ML methods, as the distance increases.

As a result of the experiments, the phenomena that cannot be explained with only the diffusion of the molecules are revealed. The molecules are affected by the initial drift of the TX, Brownian motion and gravity. Moreover, a possible trajectory of the molecules in which there are two propagation zones as the unsteady flow zone and the Brownian motion zone is given. It is analyzed that the transmitted molecules are affected by unsteady flows due to the induced air as soon as they are emitted from the TX. Furthermore, our analysis shows that the non-linear characteristic of the sensor can cause  measurement errors. The joint effect of the non-linearity of the sensor and  unsteady flows complicates the estimation of the distance at the RX side as revealed by the experimental data.


Hence, it can be concluded that a fluid dynamics perspective is needed for a precise distance and any other channel parameter estimation. It is also necessary to consider the imperfection of the TX and RX to improve parameter estimation in macroscale practical MC system models. As the future work, we plan to improve the channel parameter estimation performance by developing methods with the fluid dynamics perspective. Also, we plan to develop a method for the direction estimation of  the  TX by using multiple sensors. 
\section*{Acknowledgment}
This work was supported by the Scientific and Technological Research Council of Turkey (TUBITAK) under Grant 115E362.


\bibliography{de_fg}
\end{document}